\documentclass[twocolumn,epjc3]{svjour3}          

\RequirePackage[T1]{fontenc}

\smartqed  

\RequirePackage{graphicx}
\RequirePackage{mathptmx}      
\RequirePackage[numbers,sort&compress]{natbib}
\RequirePackage{graphicx}
\RequirePackage{flushend}
\RequirePackage[numbers,sort&compress]{natbib}

\RequirePackage{multirow}
\RequirePackage{amsmath,amssymb,amsfonts}
\RequirePackage{pifont}
\RequirePackage{algorithm}
\RequirePackage{algpseudocode}
\RequirePackage{mathrsfs}
\RequirePackage[title]{appendix}
\RequirePackage[table]{xcolor}
\RequirePackage{colortbl}
\RequirePackage{textcomp}
\RequirePackage{subcaption}
\RequirePackage{newtxtext,newtxmath}
\RequirePackage{listings}
\RequirePackage{manyfoot}
\RequirePackage{booktabs}
\usepackage{tabularx}
\usepackage{array}
\RequirePackage{array}
\RequirePackage{pifont}
\RequirePackage{graphicx}
\RequirePackage[numbers,sort&compress]{natbib}
\RequirePackage[colorlinks,citecolor=blue,urlcolor=blue,linkcolor=blue]{hyperref}
\usepackage{microtype}
\usepackage{xurl}

\newcommand{\cmark}{\checkmark}
\newcommand{\xmark}{\ding{55}}

\RequirePackage[colorlinks,citecolor=blue,urlcolor=blue,linkcolor=blue]{hyperref}

\journalname{Eur. Phys. J. C}
\usepackage{lineno} 

\begin{document}
\title{NeuroGraph: An AI Graph-Driven Neuro-Symbolic Framework for Explainable Threat Reasoning in Advanced Manufacturing}

\author{
Padmeswari Nandiya\thanksref{addr1, em1}
\and
Ahmad Mohsin\thanksref{addr1,cor1}
\and
Iqbal H.\ Sarker\thanksref{addr1, em2}
\and
Ahmed Ibrahim\thanksref{addr1, em3}
\and
Helge Janicke\thanksref{addr1, em4}
}

\institute{
School of Science (Computing \& Security Discipline),
Edith Cowan University, Perth, WA 6027, Australia\\
\label{addr1}
}

\thankstext{cor1}{
Corresponding author: a.mohsin@ecu.edu.au
}
\thankstext{em1}{e-mail: p.nandiya@ecu.edu.au} 
\thankstext{em2}{e-mail: m.sarker@ecu.edu.au} 
\thankstext{em3}{e-mail: ahmed.ibrahim@ecu.edu.au} 
\thankstext{em4}{e-mail: h.janicke@ecu.edu.au}

\institute{
School of Science (Computing \& Security Discipline),
Edith Cowan University, Perth, WA 6027, Australia
\label{addr1}
}

\date{Received: date / Accepted: date}

\maketitle

\begin{abstract}
The growing complexity of cyber-physical attack surfaces in advanced manufacturing environments has made reliable cyber threat intelligence analysis increasingly challenging. While large language models and retrieval-augmented generation have enhanced cyber threat intelligence capabilities, text-based approaches remain prone to hallucinations and lack structured reasoning over interconnected threats. Graph-based retrieval-augmented generation partially addresses this limitation but often fails to support ontology-consistent multi-hop reasoning and transparent evidence tracing across heterogeneous cybersecurity knowledge. This paper proposes a graph-grounded neuro-symbolic framework that integrates ontology-aware symbolic query generation, knowledge graph retrieval, and neural language generation to enable accurate and explainable threat analysis across information technology and operational technology domains. The proposed framework employs a dual-large language model architecture in which the first model translates natural-language queries into executable Cypher statements for symbolic graph retrieval, while the second generates responses exclusively from the retrieved graph evidence. Experimental evaluation on publicly available cyber threat intelligence benchmarks demonstrates consistent improvements over the published baseline in reasoning accuracy while reducing hallucinations, improving multi-hop reasoning, and increasing robustness against adversarial perturbations. Runtime and explainability analyses further show that the framework maintains interactive inference performance while exposing graph-grounded reasoning artifacts that enable analysts to inspect and verify each stage of the reasoning process. These results demonstrate the potential of graph-grounded neuro-symbolic reasoning to provide scalable, interpretable, and reliable cyber threat intelligence for next-generation Industry 5.0 environments.
\end{abstract}


\section{Introduction}
\label{sec:introduction}
Industry~5.0 represents the emerging evolution of advanced manufacturing, placing emphasis on Human-centric collaboration between domain experts and intelligent systems~\cite{Nahavandi2019Industry5}. In this context, Industrial Control Systems (ICS) constitute a foundational component, as they govern and automate processes across modern industrial environments. The increasing integration of intelligent and connected technologies within ICS amplifies the importance of AI-driven assistants capable of supporting cyber resilience and informed decision-making \cite{ srinivasan2025threatbasedsecuritycontrolsprotect}.

ICS are extensively deployed across critical infrastructure sectors, including energy, manufacturing, and essential utilities, where their secure and continuous operation is of strategic importance~\cite{bhamare2020cybersecurityindustrialcontrolsystems}. However, the ongoing digitalization and increasing connectivity of these systems have substantially expanded their attack surface, resulting in heightened exposure to cyber threats~\cite{Nankya2023ICS}. This growing vulnerability has been demonstrated by several high-profile cyber incidents. Notably, the Stuxnet attack in 2010 and the Triton attack in 2017 illustrated the capability of adversaries to manipulate and disrupt physical industrial processes through cyber means~\cite{FIROOZJAEI2022100487, SAINI2026110967}. More recently, the 2021 breach of the Oldsmar water treatment facility revealed that even small-scale municipal ICS deployments remain susceptible to cyber compromise~\cite{Homaei_2026}. Collectively, these incidents demonstrate that cyber attacks targeting ICS extend beyond digital infrastructures to produce tangible physical consequences, emphasizing the need for intelligent security mechanisms capable of assisting analysts in understanding and mitigating complex multi-stage cyber-physical attack scenarios.

Recent advances in Large Language Models (LLMs) have significantly improved cyber threat intelligence (CTI) analysis through Retrieval-Augmented Generation (RAG), enabling language models to access external cybersecurity knowledge during inference~\cite{borah2025adaptinglargelanguagemodels, gusarov2025multiagentgraphragtexttocypherframework, cyberally2025, mohsin2025humanai}. While promising, conventional text-based RAG systems remain susceptible to hallucinations and often struggle to perform structured reasoning over interconnected cyber-physical environments where attack paths span information technology (IT), operational technology (OT), and physical processes~\cite{saleh-etal-2024-sg, fang2024, chen2023benchmarkinglargelanguagemodels}. To address these limitations, recent studies have proposed Graph Retrieval-Augmented Generation (Graph-RAG), which leverages knowledge graphs to provide structured and relational reasoning across assets, vulnerabilities, weaknesses, and attack patterns~\cite{zhang2025surveygraphretrievalaugmentedgeneration}. Although Graph-RAG represents a significant advancement over text-only retrieval, existing approaches remain limited in supporting the heterogeneous, multi-hop reasoning required in Industry~5.0 cyber-physical environments.

Several practical challenges continue to hinder the deployment of Graph-RAG in industrial cybersecurity. First, cybersecurity knowledge is inherently heterogeneous, requiring the integration of diverse intelligence sources such as CVE, CWE, MITRE ATT\&CK, CAPEC, asset inventories, and incident reports across both IT and OT domains~\cite{sikos2023cybersecurity,mohsin2025humanai,electronics15030552}. Existing cybersecurity ontologies frequently model only subsets of this ecosystem, leaving important cross-domain relationships insufficiently represented~\cite{lourenscc,ZDEMIRSNMEZ2022102938}. Second, even when comprehensive knowledge graphs are available, cybersecurity analysts rarely interact directly with graph query languages, while existing graph-based retrieval systems provide limited support for transparent multi-hop reasoning over complex industrial environments~\cite{LIU2022108870}. Finally, although Graph-RAG enables language models to retrieve structured graph knowledge, many existing systems still provide limited reasoning transparency, making it difficult for analysts to inspect intermediate reasoning steps and verify how conclusions are derived~\cite{moghaddam2025explainableknowledgegraphretrievalaugmented,articlesssss}. These challenges highlight the need for a graph-grounded reasoning framework capable of combining structured symbolic reasoning with the flexibility of neural language models while maintaining explainability for cyber-physical threat analysis.
These challenges motivate the development of a graph-grounded neuro-symbolic framework that combines the transparency of symbolic reasoning with the flexibility of neural language models to enable explainable and reliable cyber threat analysis across interconnected IT and OT environments. Accordingly, this paper proposes \textbf{NeuroGraph}, a graph-grounded neuro-symbolic framework for threat reasoning in advanced manufacturing systems, implemented through the \textbf{GRICS} (Graph-Integrated Retrieval for Industry-Centric Security) architecture. GRICS combines a cyber-physical knowledge graph with a dual-LLM architecture. The first LLM translates analyst queries into ontology-aware Cypher queries for symbolic knowledge retrieval, while the second LLM generates natural-language responses using only the retrieved graph evidence. By separating symbolic retrieval from language generation, GRICS provides graph-grounded reasoning that improves transparency, supports multi-hop inference across heterogeneous cyber-physical entities, and reduces unsupported model generation.

GRICS is built upon the BRIDG-ICS ontology developed in the previous work~\cite{nandiya2025bridgicsaigroundedknowledgegraphs}, which models cyber-physical assets, vulnerabilities, weaknesses, attack techniques, and adversarial behaviours within a unified industrial knowledge graph. Leveraging this ontology, GRICS enables explainable graph-grounded multi-hop reasoning across interconnected IT and OT environments. The methodological novelty of GRICS lies not merely in combining knowledge graphs with large language models, but in the way its neuro-symbolic architecture couples neural and symbolic components during retrieval and reasoning. Unlike Graph-RAG approaches that primarily rely on embedding similarity, graph expansion, clustering, or learned graph representations to construct context, GRICS adopts symbolic Cypher execution as the primary evidence-retrieval mechanism. The language model first translates an analyst query into an ontology-constrained executable program, and only evidence returned through explicit graph execution is admitted to downstream answer generation. Embedding-based retrieval is activated only when the initial symbolic query fails and serves solely to identify a graph anchor from which an ontology-compliant Cypher query is regenerated. Consequently, neural retrieval assists symbolic reasoning rather than replacing it, enabling deterministic graph traversal, ontology-consistent multi-hop reasoning, and explicit evidence traceability across interconnected IT and OT cybersecurity entities.

The proposed framework assumes the availability of a curated cybersecurity knowledge graph derived from the BRIDG-ICS ontology and focuses on graph-grounded threat reasoning rather than ontology construction or maintenance. Consequently, the current study does not address automated ontology evolution, continuous cyber threat intelligence ingestion, or real-time knowledge graph synchronization. These assumptions, together with the scope and current limitations of the framework, are discussed in detail in Section~IV. The primary contributions of GRICS are summarised as follows:
nandiya2025bridgicsaigroundedknowledgegraphs
\begin{itemize}
    \item \textbf{Graph-Grounded Threat Reasoning for Smart Manufacturing.}
    We propose a graph-grounded retrieval framework that combines knowledge graph retrieval with LLM reasoning to enable multi-hop inference across assets, vulnerabilities, weaknesses, and attack techniques for cyber-physical threat analysis.

    \item \textbf{Symbolic-First Neuro-Symbolic Retrieval.} We introduce a symbolic-first retrieval architecture in which ontology-constrained Cypher execution constitutes the primary reasoning pathway, while embedding retrieval is restricted to recovery of graph anchors following symbolic-query failure. Unlike hybrid retrieval schemes in which neural and symbolic evidence are independently combined, GRICS requires recovered anchors to be converted back into executable ontology-compliant Cypher queries before evidence is admitted to downstream reasoning.

    \item \textbf{Explainable NeuroGraph Reasoning.}
    We develop a transparent reasoning workflow that exposes the generated Cypher query, retrieved graph evidence, and grounded natural-language response, enabling analysts to inspect and verify each stage of the reasoning process.
\end{itemize}

The remainder of this paper is organized as follows. Section~\ref{sec2} introduces the technical preliminaries underlying graph-grounded cyber threat reasoning. Section~\ref{sec3} reviews the related research and identifies the existing research gaps. Section~\ref{sec:problem} formulates the research problem and outlines the assumptions and scope of the proposed framework. Section~\ref{sec5}  presents the proposed graph-grounded neuro-symbolic framework. Section~\ref{sec:evaluation} describes the experimental evaluation and discusses explainability and robustness analyses. Finally, Section~\ref{sec:discussions} concludes the paper and outlines future research directions.

\section{Preliminaries}
\label{sec2}
\subsection{Cybersecurity Knowledge Graphs}
In Industry 5.0 environments, Industrial Control Systems (ICS) face increasingly complex cyber threats due to interconnected and AI-enabled infrastructures~\cite{Nahavandi2019Industry5,Rejeb2025Industry5}. Knowledge graphs (KGs) provide a structured representation that integrates diverse cybersecurity data across IT and OT domains, enabling a unified and machine-interpretable view of system entities and their relationships~\cite{liu2022, sikos2023cybersecurity,nandiya2025bridgicsaigroundedknowledgegraphs}.
These relationships support multi-hop reasoning over interconnected components, facilitating the analysis of attack dependencies and system-level risks~\cite{ics_sec}.
Formally, a cybersecurity knowledge graph is represented by the graph model shown in Equation~(\ref{eq:graph-definition}):

\begin{equation}
\mathcal{G} = (\mathcal{V}, \mathcal{E}, \mathcal{R})
\label{eq:graph-definition}
\end{equation}

\subsection{Graph Retrieval-Augmented Generation}

Retrieval-Augmented Generation (RAG) has the potential to enhance LLMs by incorporating external knowledge during inference, thereby improving factual grounding~\cite{lewis2021, hu2025rag,fan2024}. However, it relies on unstructured text, limiting its ability to capture relationships between cybersecurity entities.
Graph-based RAG (Graph-RAG) addresses this limitation by retrieving structured subgraphs $R \subseteq G$ from a knowledge graph $G$, enabling multi-hop reasoning across interconnected entities~\cite{cai2025simgrag, zhu2025, hu2025grag}. Graph-RAG enables LLMs to reason over structured, multi-hop knowledge by integrating symbolic graph traversal with embedding-based similarity search.

\subsection{Neuro-Symbolic Reasoning}
Symbolic–semantic reasoning maps natural language inputs to executable structured representations over knowledge graphs, enabling precise, explainable inference. Given an input $q \in X^*$, the goal is to derive $c \in C$ such that executing it on graph $G$ yields relevant results, $R = \mathrm{exec}(c, G)$. This combines semantic interpretation with symbolic graph operations to support multi-hop reasoning over interconnected cybersecurity entities, providing traceable, context-aware threat analysis in cyber–physical environments.

\subsection{Cyber-Physical Threat Modelling}

Cyber-physical threat modelling represents security events as interconnected relationships among digital assets, operational components, vulnerabilities, weaknesses, attack patterns, adversarial techniques, and physical processes~\cite{BARRERE2023103348}. In industrial environments, security incidents rarely remain confined to a single software component or network layer. Instead, compromise can propagate across information technology (IT) and operational technology (OT) domains, affecting industrial assets, control processes, and potentially physical operations~\cite{srinivasan2025threatbasedsecuritycontrolsprotect}.

A cyber-physical attack can therefore be viewed as a sequence of dependent security events rather than an isolated vulnerability~\cite{mohsin2025humanai}. For example, exploitation of a software vulnerability may expose an underlying weakness, enable a particular attack pattern, correspond to a known adversarial technique, and subsequently affect an operational asset or process~\cite{Rahman_2024}. Representing these dependencies explicitly supports the analysis of multi-stage attack paths. Such analysis requires consideration of both semantic relationships among cybersecurity concepts and operational relationships among assets, systems, and networked components~\cite{nandiya2025bridgicsaigroundedknowledgegraphs,BARRERE2020102471}.

These structured threat representations are particularly important in Industry~5.0 environments, where increased connectivity among intelligent systems, industrial assets, and human operators creates complex interdependencies~\cite{LIU2023103131}. Cyber-physical threat modelling therefore provides a conceptual foundation for analyzing multi-hop attack scenarios, correlating heterogeneous cybersecurity information, and supporting explainable security decision-making across interconnected IT and OT infrastructures.

\section{Related Work}
\label{sec3}
This section reviews the literature most relevant to graph-grounded cyber threat intelligence and positions the proposed GRICS framework within the current state of the art. The reviewed studies were selected to represent the principal research directions relevant to this work, including cybersecurity knowledge graphs, retrieval-augmented generation, Graph-RAG, neuro-symbolic reasoning, and explainable artificial intelligence. Priority was given to recent studies (2022--2025) to reflect the rapid development of Graph-RAG and large language models in cybersecurity, while seminal works were retained where necessary to provide foundational context. Representative studies were further selected based on their relevance to industrial control systems, cyber-physical systems, and cyber threat intelligence.
Based on these criteria, the literature is organized into four themes: (i) retrieval-augmented generation and Graph-RAG, (ii) neuro-symbolic retrieval and reasoning, (iii) threat reasoning in cybersecurity, and (iv) ontological modelling for cybersecurity. This structure allows GRICS’s research gaps to be systematically identified.

\subsection{Retrieval-Augmented Generation for Cyber Threat Intelligence}
\textbf{RAG for LLM Learning.}
Large language models (LLMs) excel at natural-language understanding and generation but rely on parametric knowledge, limiting factual reliability and adaptability to evolving domains~\cite{zheng2024reliablellmsknowledgebases}. Retrieval-Augmented Generation (RAG) addresses these issues by grounding outputs in external knowledge, improving factual accuracy and timeliness~\cite{lewis2021, karpukhin2020densepassageretrievalopendomain}. However, most RAG frameworks retrieve from unstructured text~\cite{Wang2025RAGKnowledgeGraph, zhang2024llmm}, which offers limited structure and is insufficient for multi-step reasoning over complex relations~\cite{guu2020realmretrievalaugmentedlanguagemodel}.

Recent advances have sought to overcome these limitations by integrating structured graph-based knowledge into retrieval-augmented reasoning~\cite{zhang2025surveygraphretrievalaugmentedgeneration}. Graph representations explicitly encode entities and relations, enabling richer contextual grounding and more effective multi-step inference. This has led to the emergence of Graph-RAG frameworks, which extend traditional RAG by leveraging structured knowledge to improve retrieval quality and reasoning interpretability. Existing Graph-RAG frameworks improve structured retrieval but remain limited in supporting deterministic graph traversal, ontology-aware reasoning, and transparent multi-hop inference over heterogeneous cyber-physical knowledge. These limitations motivate the graph-grounded neuro-symbolic design adopted by GRICS.

\textbf{Graph-RAG Methods for Cyber Threat Intelligence.}
Following the emergence of Graph-RAG frameworks, existing approaches differ in how graph structures are incorporated into retrieval and reasoning processes. These systems can be broadly characterized according to their underlying retrieval strategies, which directly influence their ability to support multi-hop reasoning, scalability, and interpretability.

The \textbf{\textit{first category}} consists of \textit{embedding-based retrieval methods}, which map entities and relations into continuous vector spaces and perform semantic similarity search~\cite{grover2016node2vec, xie2023lambdakglibrarypretrainedlanguage, chen2025kgbilmknowledgegraphembedding}. These approaches support efficient large-scale retrieval and are widely adopted in RAG pipelines. However, because they rely on approximate matching, they do not explicitly enforce structural constraints and may therefore retrieve semantically relevant but causally invalid connections.
A \textbf{\textit{second category}} addresses this limitation through \textit{subgraph and multi-hop expansion}, where reasoning is performed by iteratively exploring connected graph neighborhoods. Systems such as Think-on-Graph~2.0~\cite{ma2025think} explicitly guide LLM reasoning through graph traversal, while KGMP~\cite{yang2025kgmp} and DRKG~\cite{app15126722} introduce structured multi-hop reasoning strategies to improve interpretability. Similarly, SimGRAG~\cite{cai2025simgrag} and KG2RAG~\cite{zhu2025} retrieve subgraphs to capture broader relational context. Although these approaches improve multi-step reasoning, unconstrained expansion in densely connected graphs may introduce irrelevant or redundant paths.

The \textbf{\textit{third category}} focuses on \textit{hierarchical and structured retrieval methods}, which seek to improve scalability and support reasoning over larger or longer-context knowledge spaces. RAPTOR~\cite{sarthi2024raptor}, for instance, organizes information into tree-like structures for recursive retrieval, while GraphRAG~\cite{han2024graphrag} employs clustering and summarisation to improve knowledge organisation. Recent frameworks such as GRAG~\cite{hu2025grag} further extend graph-centric retrieval to improve flexibility and retrieval effectiveness.
Finally, the \textbf{\textit{fourth category}} comprises \textit{neural graph reasoning approaches}, which use graph neural networks and hybrid architectures to learn complex dependencies directly from graph structure~\cite{wu2020gnn, Cao_2025, mavromatis2024gnnraggraphneuralretrieval}. These methods provide strong representation-learning capabilities and can capture complex relational patterns. However, their reasoning processes are often encoded implicitly within model parameters, which can limit transparency and make intermediate reasoning paths more difficult to inspect.
Despite these advances, most Graph-RAG systems still provide limited support for explicit, constraint-driven reasoning. As shown in Table~\ref{tab:rag_comparison}, symbolic querying remains underused, restricting deterministic and interpretable multi-hop inference. This limitation is particularly important in cybersecurity, where general-purpose Graph-RAG systems often lack cyber-physical semantics, attacker-behaviour models, and explicit IT/OT integration required for \textbf{Industry~5.0} threat intelligence. Cybersecurity-oriented RAG systems such as MoRSE~\cite{simoni2024} and ProveRAG~\cite{fayyazi2025} primarily rely on unstructured text retrieval, while graph-based approaches such as CyKG-RAG~\cite{cykg} and GraphRAG under Fire~\cite{liang2025graphrag} improve structured retrieval but remain limited in representing complex, multi-stage cyber--physical dependencies.

\subsection{Neuro-Symbolic Reasoning in Graph-RAG}

Neuro-symbolic approaches aim to integrate data-driven learning with 
structured reasoning over knowledge graphs, combining the flexibility 
of neural models with the consistency of symbolic representations~\cite{liu2024neuralsymbolicreasoningknowledgegraphs, DeLong_2025}. In 
Graph-RAG systems, this paradigm enables the use of learned embeddings 
for retrieval while leveraging graph structures to provide contextual 
grounding for reasoning~\cite{ZHANG202114}.

In practice, many existing approaches emphasize neural retrieval mechanisms, 
where embeddings or learned representations guide the selection of relevant 
nodes and subgraphs~\cite{NEURIPS2023_5965f3a7}. While effective for scalability and generalization, 
such approaches do not explicitly enforce logical constraints, and reasoning 
is often guided by similarity rather than structural validity~\cite{10.1145/3686806}. As a result, 
multi-hop inference may include semantically relevant but structurally 
inconsistent relationships, particularly in complex domains~\cite{LIU2024127571}.
Symbolic querying (e.g., Cypher, SPARQL) provides an alternative by enabling 
deterministic and constraint-driven traversal over knowledge graphs~\cite{angles2018graph, purkayastha2021knowledgegraphquestionanswering, zhao2023s2ctransbuildingbridgesparql}. 
By enforcing explicit structural conditions, symbolic methods ensure that 
all inferred relationships adhere to valid graph connections, supporting 
precise and interpretable reasoning.
Existing neuro-symbolic Graph-RAG systems, however, rarely integrate explicit symbolic query generation and neural retrieval within a unified reasoning pipeline. As a result, deterministic graph traversal, ontology-aware reasoning, and transparent evidence tracing remain underexplored, particularly in cyber-physical threat intelligence.

Recent graph-grounded reasoning approaches have increasingly moved beyond static retrieval toward multi-strategy, agentic, and neuro-symbolic architectures. BYOKG-RAG combines LLM-generated graph artifacts, including candidate entities, reasoning paths, and OpenCypher queries, with specialized graph-retrieval tools and iterative refinement over heterogeneous knowledge graphs~\cite{mavromatis2025byokgrag}. SymAgent adopts an agentic neuro-symbolic architecture in which an Agent-Planner extracts symbolic reasoning structures from the knowledge graph and an Agent-Executor dynamically invokes external and graph-based tools to address complex reasoning tasks and incomplete graph knowledge~\cite{liu2025symagent}. In parallel, TUNSR investigates a unified neuro-symbolic reasoning framework that combines neural representations with symbolic first-order logic reasoning over dynamically constructed reasoning graphs~\cite{lin2025unified}. These approaches demonstrate important advances toward adaptive graph interaction, iterative tool use, and tighter integration between neural and symbolic reasoning.

GRICS differs from these recent approaches in both its reasoning objective and the functional roles assigned to its neural and symbolic components. BYOKG-RAG employs multiple graph-retrieval strategies to improve knowledge-graph question answering, whereas GRICS constrains accepted evidence to ontology-compliant Cypher execution over the BRIDG-ICS cybersecurity ontology. Compared with agentic architectures such as SymAgent, which rely on iterative planning and dynamic tool invocation, GRICS adopts a more constrained dual-LLM reasoning pipeline in which the first LLM generates executable symbolic queries and the second synthesises responses only after graph evidence has been retrieved. Relative to broader neuro-symbolic frameworks such as TUNSR, GRICS is specifically designed for explainable cybersecurity reasoning across vulnerability, weakness, attack-pattern, MITRE ATT\&CK, and industrial-asset relationships.

A key distinction is that semantic retrieval in GRICS cannot independently determine the final reasoning evidence. Embedding-based retrieval is used to recover candidate graph anchors when symbolic retrieval fails, but the recovered candidate must subsequently pass through ontology-compliant Cypher generation and graph execution before being supplied to the Answer-LLM. This establishes an explicit verification boundary between semantic interpretation and factual retrieval. Consequently, the principal advantage of GRICS is not unrestricted reasoning flexibility, but the combination of neural adaptability with deterministic graph traversal, ontology consistency, and end-to-end evidence traceability required for cybersecurity decision support.


\begin{table*}[!t]
\centering
\caption{Comparison of representative RAG and Graph-RAG approaches highlighting gaps in reasoning, constraints, and retrieval mechanisms.}
\label{tab:rag_comparison}
\resizebox{\textwidth}{!}{
\begin{tabular}{p{3.6cm}p{2cm}p{2.1cm}p{2.1cm}p{1.9cm}p{2.4cm}p{1.6cm}p{1.8cm}}
\hline
\textbf{Work (Year)}
& \textbf{Symbolic Reasoning}
& \textbf{Multi-hop Consistency}
& \textbf{Constraint-based Retrieval}
& \textbf{Embedding-based Retrieval}
& \textbf{Retrieval Type}
& \textbf{KG}
& \textbf{LLM Fine-Tuning} \\
\hline

GraphRAG (2025)~\cite{han2024graphrag}
& \xmark & \xmark & \xmark & \cmark & Clustering / Subgraph & \cmark & \xmark \\

KG2RAG (2025)~\cite{zhu2025}
& \xmark & \xmark & \xmark & \cmark & Graph-guided Retrieval & \cmark & \xmark \\

GRAG (2025)~\cite{hu2025grag}
& \xmark & \xmark & \xmark & \cmark & Graph Retrieval & \cmark & \xmark \\

GNN-RAG (2024)~\cite{mavromatis2024gnnraggraphneuralretrieval}
& \xmark & \xmark & \xmark & \cmark & Neural Graph Reasoning & \cmark & \xmark \\

MoRSE (2024)~\cite{simoni2024}
& \xmark & \xmark & \xmark & \cmark & Textual Retrieval & \xmark & \xmark \\

CyKG-RAG (2024)~\cite{cykg}
& \xmark & $\triangle$ & \xmark & \cmark & KG-based Retrieval & \cmark & \xmark \\

\hline

\rowcolor{gray!15}
\textbf{GRICS (Ours)}
& \cmark & \cmark & \cmark & \cmark
& Neural + Symbolic (Cypher-based)
& \cmark & \cmark \\

\hline
\end{tabular}
}
\end{table*}
\subsection{Threat Reasoning in Cybersecurity}

Threat reasoning has become an increasingly important research direction for supporting cyber threat intelligence (CTI) analysis, enabling systems to infer relationships among vulnerabilities, attack techniques, threat actors, and defensive actions rather than performing isolated information retrieval. Recent advances have leveraged large language models and knowledge graphs to improve structured reasoning over cybersecurity knowledge.

Fieblinger \textit{et al.}~\cite{fieblinger2024actionablecyberthreatintelligence} combine knowledge graphs with large language models to transform unstructured CTI reports into actionable threat intelligence, facilitating contextual analysis and information extraction. Wu \textit{et al.}~\cite{wu2025kgvintegratinglargelanguage} further extend this direction by integrating knowledge graphs with large language models for cyber threat intelligence credibility assessment, demonstrating how structured semantic relationships can improve the reliability of CTI. More recently, CTI-Thinker~\cite{Yang2026} incorporates ATT\&CK-aligned knowledge graphs within a Graph-RAG framework to support attack intent inference and CTI question answering.

These studies demonstrate that structured knowledge substantially improves cybersecurity reasoning compared with language-model-only approaches. However, existing methods primarily rely on \textbf{\textit{GraphRAG}} or embedding-based retrieval over cybersecurity knowledge graphs. Consequently, the reasoning process remains largely implicit within the language model and provides limited support for deterministic graph traversal or explicit verification of intermediate reasoning steps. In contrast, this work investigates graph-grounded neuro-symbolic reasoning through ontology-aware Cypher generation and \textbf{\textit{symbolic graph}} execution, enabling transparent multi-hop reasoning in which every generated response can be traced to explicit graph evidence.

\vspace{-1.3em}
\subsection{Ontological Modelling in Cybersecurity}

Ontological modelling has been central to cybersecurity research, enabling structured and machine-interpretable representations of threat intelligence, vulnerabilities, and attack behaviours. Early efforts such as \textit{MITRE ATT\&CK} and \textit{CAPEC} ontologies~\cite{mitreattack2020, capec2019}, along with frameworks like \textit{STIX}~\cite{stix2017}, \textit{CVE}, and \textit{CWE}, established standardised representations to support knowledge sharing and analysis.
Recent work extends these foundations into Cyber Knowledge Graphs (CKGs), integrating heterogeneous data sources and capturing relationships among entities for contextual reasoning~\cite{uco2019, octave2020, cti_kg2022, cybonto2022, informatics12030100}. However, these approaches often emphasize structural completeness over adaptive reasoning, limiting their ability to support dynamic, multi-hop analysis in cyber--physical environments.

As highlighted in Table~\ref{tab:rag_comparison}, current RAG and Graph-RAG approaches do not provide integrated symbolic reasoning, cyber–physical modelling, or explainable threat analysis, which restricts their applicability in complex, safety-critical settings. These limitations are addressed by the proposed GRICS framework.

\section{Problem Statement}
\label{sec:problem}

The increasing adoption of Industry~5.0 technologies has created highly interconnected cyber-physical environments spanning information technology (IT), operational technology (OT), industrial assets, vulnerabilities, and attacker behaviours. Effective cyber threat intelligence (CTI) therefore requires reasoning across heterogeneous entities and multiple abstraction levels rather than relying on isolated document retrieval or keyword matching.

Although Retrieval-Augmented Generation (RAG) improves factual grounding through external knowledge, conventional text-based retrieval lacks explicit representations of relationships among cybersecurity entities. Recent Graph-RAG approaches incorporate knowledge graphs, but many still rely on embedding similarity, heuristic graph expansion, or summarisation, limiting deterministic traversal, ontology compliance, and transparent multi-hop reasoning. Cybersecurity analysts also typically express investigation goals in natural language rather than graph query languages, creating a need for systems that can interpret user intent while preserving ontology constraints and reasoning traceability. GRICS addresses these limitations through a unified neuro-symbolic pipeline that combines ontology-aware Cypher generation, symbolic graph retrieval, embedding-assisted query recovery, and grounded natural-language response generation. This work is developed under the following assumptions:

\begin{itemize}
    \item The cybersecurity knowledge graph is constructed from a trusted ontology and accurately represents cyber-physical entities and their relationships.
    \item Retrieved graph information is considered authoritative, while the language model is responsible only for interpreting and summarizing retrieved evidence rather than generating unsupported knowledge.
    \item Analysts interact with the framework using natural-language queries rather than directly writing graph query languages such as Cypher.
\end{itemize}

The scope of this work is limited to graph-grounded cyber threat reasoning over the BRIDG-ICS ontology. The framework does not address automatic ontology construction, knowledge graph population, or ontology evolution. Likewise, although explainability is achieved through transparent graph-grounded reasoning artifacts, formal human-subject evaluation of analyst trust and cognitive workload is beyond the scope of this study and remains an important direction for future work.

\section{Proposed GRICS Framework}
\label{sec5}

This section presents the proposed NeuroGraph-based neuro-symbolic reasoning framework, referred to as Graph-Integrated Retrieval for Industry-Centric Security (GRICS), for advanced threat reasoning in Industry~5.0 cyber--physical environments. GRICS is implemented as a domain-specific knowledge-graph-based retrieval-augmented generation (KG-RAG) framework that enables multi-hop attack-path analysis, threat-impact assessment, and mitigation-strategy generation by combining structured cyber--physical knowledge with large language model reasoning. The framework integrates three core components: (i) a cyber--physical knowledge graph derived from the BRIDG-ICS ontology that unifies information-technology and operational-technology assets, (ii) a dual-LLM pipeline that separates ontology-aware symbolic retrieval from answer synthesis, and (iii) a threat-centric KG-RAG question-answering dataset designed to support multi-hop reasoning and attack-scenario analysis.

\begin{figure*}[t]
    \centering
    \includegraphics[width=0.85\linewidth]{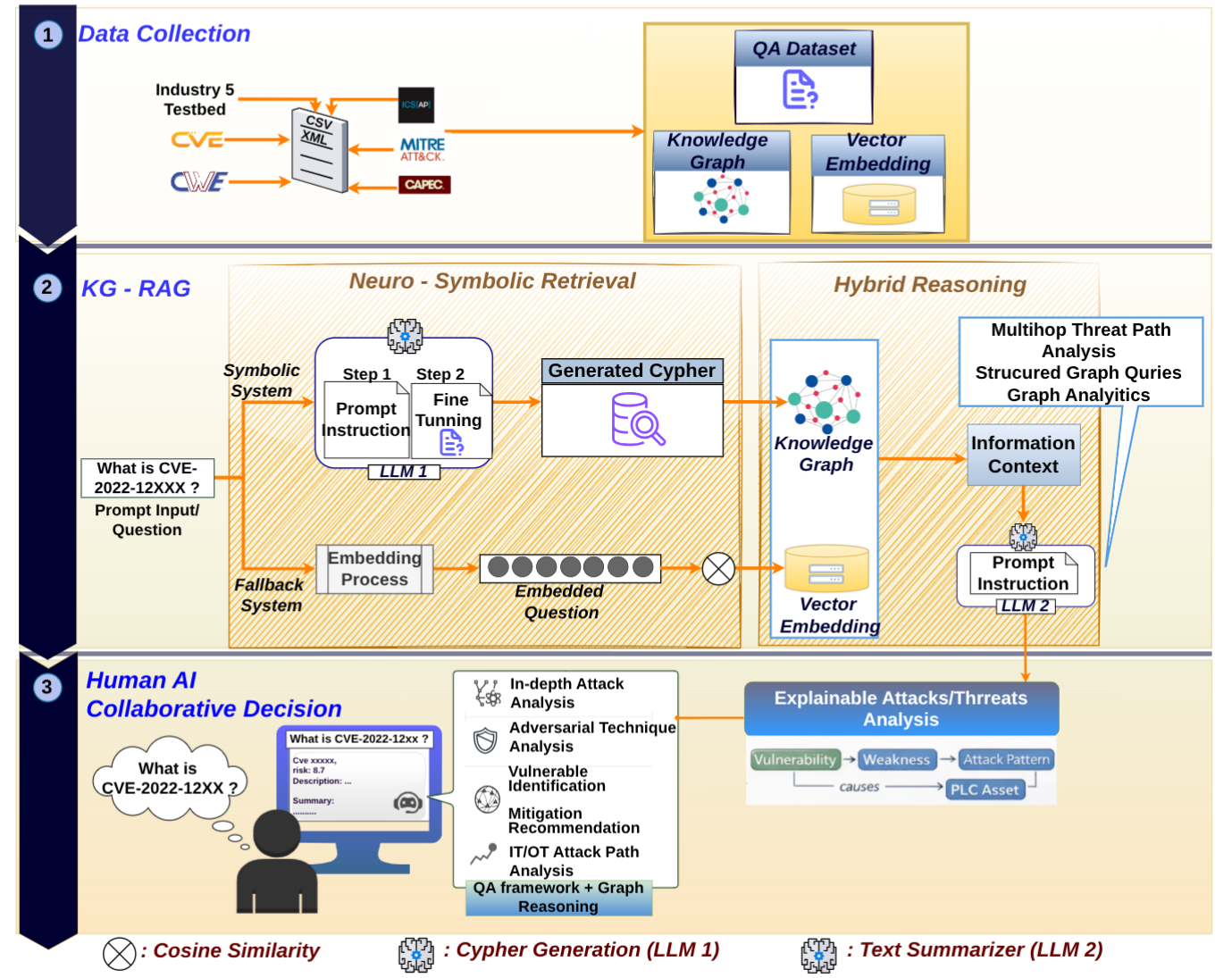}
    \caption{GRICS KG-RAG architecture and dual LLM reasoning pipeline.}
    \label{fig:grics_framework}
\end{figure*}

The design of GRICS is informed by recent advances in knowledge-graph-guided RAG systems, which demonstrate that structured knowledge can improve retrieval precision, interpretability, and reasoning robustness. GRICS extends these approaches through neuro-symbolic graph reasoning by combining large language model interpretation with ontology-constrained Cypher generation and explicit knowledge-graph traversal. In contrast to general-domain KG-RAG approaches, GRICS is tailored to Industrial Control Systems security, with a focus on interpretable attack-path reasoning and cyber--physical dependency modelling. An overview of the GRICS architecture is shown in Figure~\ref{fig:grics_framework}. The implementation details and source code are publicly available\footnote{\url{https://github.com/ahmadspm/Resellient-Industry-5.0--kG-Digital-Twins/tree/main}}.

\subsection{Data Collection}

Industrial cyber-physical environments are highly interconnected and often exhibit cascading security risks, where vulnerabilities in network infrastructure or security controls can propagate across multiple factory components. Capturing such interactions requires integrating heterogeneous cybersecurity information from both cyber and operational domains.

The datasets used in this study are collected from multiple authoritative cybersecurity repositories, including Common Vulnerabilities and Exposures (CVE), Common Platform Enumeration (CPE), Common Weakness Enumeration (CWE), CAPEC, MITRE ATT\&CK, and industrial asset information obtained from real-world Industrial Internet of Things (IIoT) and Operational Technology (OT) testbeds. These heterogeneous sources are transformed into node and edge CSV files and represented using the labelled property graph (LPG) model, in which cybersecurity entities are encoded as labelled nodes, relationships are represented as typed edges, and descriptive attributes are stored as node or relationship properties. The resulting LPG is implemented in Neo4j, which provides the graph database environment used for knowledge-graph construction, storage, and subsequent Cypher-based retrieval.

\subsection{Knowledge Sources and Ontology}

The imported cybersecurity datasets are organized according to the BRIDG-ICS ontology~\cite{nandiya2025bridgicsaigroundedknowledgegraphs}, which provides the semantic structure for integrating industrial assets, software products, vendors, vulnerabilities (CVE), weaknesses (CWE), attack patterns (CAPEC), MITRE ATT\&CK techniques, and operational zones. Each entity is assigned a source-specific unique identifier and mapped to the corresponding ontology class, helping align records across heterogeneous cybersecurity sources and prevent duplicate nodes during graph construction.

The ontology model uses a domain-driven design where cybersecurity concepts are modelled as classes and linked by typed relationships. Relations such as \textit{has\_Vulnerability}, \textit{Attack(Asset)}, \textit{has\_CWE}, \textit{use\_Technique}, and \textit{located\_In(Zone)} enable explainable multi-hop reasoning across cyber-physical attack paths. Relationships are instantiated only when supported by source data mappings or defined in the BRIDG-ICS ontology, preventing unsupported associations during graph construction.
In this study, a fixed snapshot of the BRIDG-ICS knowledge graph is used for all training and evaluation to ensure a consistent experimental setup. Cybersecurity records from selected sources are preprocessed offline into normalised nodes and relationships following the BRIDG-ICS schema and then imported into the graph. When the graph is updated outside evaluation, embeddings are generated for new or modified nodes to keep the embedding-based fallback retrieval mechanism consistent.

\subsubsection{Knowledge Graph Construction}
\label{sec_kg}


The knowledge-graph construction pipeline consists of four stages: source extraction, entity normalization, relationship mapping, and graph ingestion. First, records from the selected cybersecurity sources are converted into intermediate node and relationship tables. Second, entities such as CVE, CWE, CAPEC, MITRE ATT\&CK techniques, and CPE records are normalised using canonical identifiers to align heterogeneous sources and reduce duplication. Third, relationships among these entities, together with industrial assets and software components, are established according to the BRIDG-ICS schema. Finally, the resulting node and relationship tables are imported into the graph database for graph-based retrieval and reasoning. The complete BRIDG-ICS ontology schema, relationship definitions, and associated implementation resources are available in the project repository.\footnote{\url{https://github.com/ahmadspm/Industry-5.0--Intelligent-Threat-Analytics-KGs-and-LLMs}}

Based on the integrated dataset, the resulting knowledge graph is represented as $\mathcal{G} = (\mathcal{V}, \mathcal{E})$, where $\mathcal{V}$ denotes the set of cyber-physical entities and $\mathcal{E}$ represents the semantic and operational relationships among them. A principal reasoning chain represented in the graph is:

\begin{quote}
Industry~5.0 Assets $\rightarrow$ CVE $\rightarrow$ CWE $\rightarrow$ CAPEC $\rightarrow$ MITRE ATT\&CK.
\end{quote}

This structure enables multi-hop reasoning for attack-path analysis and threat attribution across interconnected cybersecurity entities. In addition to these semantic relationships, the graph models operational interactions through typed edges $(u,v)\in\mathcal{E}_{\mathrm{comm}}$ using the relation \texttt{COMMUNICATES\_WITH}. These edges are enriched with risk-related attributes, including \textit{pExploit}, \textit{riskWeight}, \textit{controlStrength}, and \textit{costAttack}, supporting quantitative assessment of vulnerability propagation.

\begin{figure}[!t]
\centering
\includegraphics[width=\linewidth]{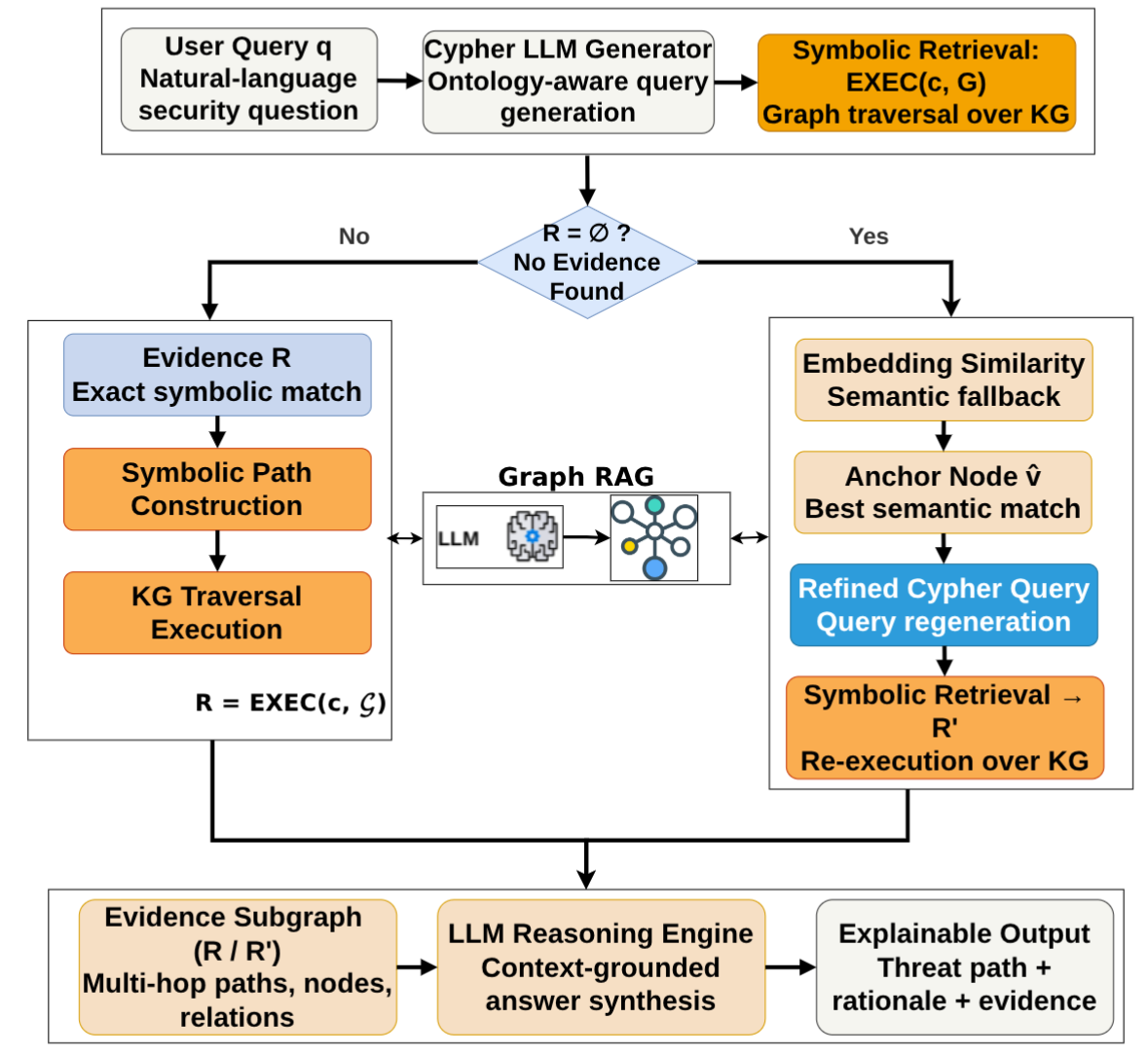}
\caption{Knowledge graph structure and neuro-symbolic retrieval workflow in GRICS.}
\label{fig-NueroRAG}
\end{figure}

\subsubsection{Vector Embedding Database}

After knowledge-graph construction, the intermediate node records stored in CSV format are transformed into dense vector representations. Each node is embedded into a 384-dimensional vector using the \texttt{all-MiniLM-L6-v2} model, which provides a balance between computational efficiency and semantic representation quality. Each row in the node CSV corresponds to a single graph entity and is converted into a fixed-length embedding representation.

The resulting embeddings are stored in a vector database to support semantic fallback retrieval. Similarity between the embedded user query and graph-node representations is computed using cosine similarity to identify semantically related candidate nodes. These candidates serve only as graph anchors for query recovery and are subsequently supplied to the Cypher-generation module to regenerate an ontology-compliant query before symbolic graph execution resumes. Given the 384-dimensional representation, each embedding requires approximately 1.5~KB of raw storage in 32-bit floating-point precision, excluding database and indexing overhead.

\vspace{-1.5em}
\subsection{KG-RAG and Reasoning Pipeline}

The KG-RAG pipeline extends the constructed knowledge graph into a query-driven reasoning system for cyber--physical threat analysis. Given a user query, the framework retrieves relevant graph-structured evidence for downstream reasoning. Unlike conventional Retrieval-Augmented Generation (RAG) approaches that rely on unstructured text retrieval, the proposed method operates directly on the knowledge graph. This enables the system to follow explicit relational paths between entities, allowing dependencies across the cyber--physical attack surface to be explored in a structured and interpretable manner.
The retrieved subgraph serves as grounded evidence for downstream reasoning, ensuring that generated responses are consistent with the underlying graph structure. This design supports reliable inference over complex attack scenarios while maintaining traceability between retrieved evidence and generated outputs. Figure~\ref{fig-NueroRAG} illustrates the graph-based retrieval and reasoning process, demonstrating how the pipeline grounds reasoning in explicit cyber--physical relationships. This design aligns with prior work showing that knowledge graph integration improves relational coherence and retrieval quality in RAG systems~\cite{10.1145/3777378,mohsin2025humanai,cyberally2025}.

\subsubsection{Information Retrieval via Symbolic \& Controlled Fallback}

The symbolic retrieval process begins by translating a natural-language cybersecurity query into an executable Cypher query. To improve the reliability of query generation, prompt engineering and domain adaptation are employed to constrain the generated queries to valid node labels, relationship types, and property names defined in the BRIDG-ICS ontology. The generated Cypher query is subsequently executed over the knowledge graph to retrieve graph-grounded evidence for downstream reasoning.

A neuro-symbolic retrieval mechanism is adopted to integrate ontology-guided symbolic reasoning with neural language understanding. Given a user query $q$, a Cypher-generating language model produces an executable query $c$ conditioned on the domain ontology. This symbolic query is treated as the \emph{primary retrieval pathway}, ensuring that all retrieved evidence is grounded in the structured semantics of the knowledge graph. Executing this query over the knowledge graph $\mathcal{G}$ yields an evidence set $R = \textsf{EXEC}(c, \mathcal{G})$, where $R$ consists of ontology-grounded nodes, edges, and multi-hop paths that provide interpretable and traceable evidence for downstream reasoning.

The Cypher-generating LLM receives a structured prompt consisting of three components: (i) a compact representation of the BRIDG-ICS ontology schema, including node labels, relationship types, and property definitions; (ii) explicit instructions requiring the model to generate only executable Cypher statements using ontology-defined node labels, relationship types, and properties, without introducing entities or relations outside the BRIDG-ICS schema; and (iii) the analyst's natural-language query. The generated Cypher query is syntactically validated before execution over the knowledge graph. This ontology-guided prompting strategy constrains symbolic retrieval to valid graph structures while reducing invalid schema references and unsupported query generation.

However, the generated query $c$ may be \emph{unusable} due to syntactic errors, execution failures, or invalid schema references. To ensure robustness, a controlled embedding-based fallback mechanism is activated only when Cypher execution fails. Let $f_{\text{emb}} : \mathcal{V} \rightarrow \mathbb{R}^d$ denote a node embedding function. The query $q$ is embedded into the same space, and cosine similarity is computed against all node embeddings. The most relevant anchor node is selected according to Equation~(\ref{eq:anchor}):

\begin{equation}
\hat{v} = \arg\max_{v' \in \mathcal{V}}
\mathrm{cosine}(f_{\text{emb}}(q),f_{\text{emb}}(v'))
\label{eq:anchor}
\end{equation}
which is then used to guide the regeneration of a valid symbolic query. This design preserves interpretability while improving robustness, aligning with recent neuro-symbolic RAG approaches~\cite{mohsin2025humanai,nandiya2025bridgicsaigroundedknowledgegraphs,cyberally2025}. Consequently, the embedding-based retrieval module does not replace symbolic graph reasoning. Instead, it serves solely as a recovery mechanism that re-establishes graph-grounded retrieval when symbolic Cypher execution fails, ensuring that all downstream reasoning remains grounded in explicit knowledge graph traversal.

\subsubsection{Hybrid Information Retrieval and Implementation}

The proposed retrieval framework operationalizes the above neuro-symbolic mechanism within a practical system setting. While symbolic Cypher queries provide precise and interpretable retrieval, real-world queries often vary in structure, length, and semantic focus. In particular, CVE nodes contain heterogeneous attributes (e.g., identifier, risk score, description, Common Vulnerability Scoring System (CVSS) metrics), and user queries may target any subset of these fields, making consistent symbolic query generation challenging. In such cases, the embedding-based fallback mechanism improves retrieval robustness by capturing semantic intent. This is particularly effective for long or descriptive queries that lack explicit identifiers. For example, a query such as \emph{“What type of vulnerability allows attackers to inject malicious scripts into web pages?”} can be semantically mapped to nodes associated with cross-site scripting (e.g., CVE-2021-XXX), which then serve as anchors for symbolic query refinement and subsequent multi-hop graph traversal. 

Once the most relevant anchor node is identified through embedding similarity, it is not returned directly as the final answer. Instead, the original user query together with the recovered graph anchor are supplied back to the Cypher generation module, which regenerates an ontology-compliant Cypher query grounded on the identified entity while preserving the user's original reasoning intent. The regenerated Cypher query is subsequently executed over the knowledge graph using the same symbolic query execution pipeline as the primary reasoning process, enabling both one-hop and multi-hop traversal across connected cybersecurity entities. This traversal retrieves comprehensive graph-grounded evidence, including node attributes (e.g., descriptions, severity scores, and CVSS metrics) as well as connected entities such as weaknesses, attack techniques, and mitigations. The resulting subgraph serves as structured evidence for downstream answer generation, enabling explainable analysis of vulnerabilities, attack paths, vulnerability propagation, ATT\&CK technique attribution, and mitigation recommendations. Consequently, the embedding module functions solely as a recovery mechanism for identifying the initial graph anchor, while all subsequent reasoning remains grounded in explicit symbolic graph traversal.

\smallskip
\noindent
\textbf{Cypher Prompt Engineering.}
To improve retrieval robustness, the Cypher-generating LLM produces up to three candidate queries for each analyst request. Duplicate and syntactically invalid queries are discarded before execution over the labelled property graph, where cybersecurity entities are represented as labelled nodes, relationships as typed edges, and relevant attributes as graph properties. For lengthy or semantically complex analyst queries, the input is decomposed into multiple semantic segments using \texttt{CONTAINS} filters, allowing relevant ontology concepts to be queried independently while remaining within the context window of the language model. Fine-tuning further improves the generation of ontology-compliant Cypher queries, particularly for multi-hop reasoning involving heterogeneous cybersecurity entities and relationships. Details of the fine-tuning procedure are presented in Section~\ref{sec:fine}. The retrieval workflow is summarised in Algorithm~\ref{alg:kgrag-retrieval}, while the prompt segmentation strategy is described in Algorithm~\ref{alg:prompt-cypher}.

\smallskip
\noindent
\textbf{Design Considerations.}
The hybrid framework is informed by three key considerations: (i) \emph{token limitations}, necessitating compact ontology representations; (ii) \emph{query robustness}, where embedding-based fallback compensates for failures in symbolic query generation; and (iii) \emph{accuracy--efficiency trade-off}, balancing precise graph-based retrieval with flexible semantic matching.

\subsubsection{Two-Stage Reasoning}

The reasoning process follows a two-stage KG-RAG pipeline that separates \textbf{\textit{graph-based retrieval from answer generation}}. In the first stage, the neuro-symbolic retrieval module extracts a relevant subgraph from the knowledge graph, producing an evidence set $R$ grounded in explicit cyber--physical relationships.

In the second stage, the retrieved subgraph, including graph paths, node attributes, and relationship information, is provided as structured context to a second language model that performs only response synthesis. Unlike the Cypher-generating LLM, this model is not fine-tuned and receives an explicit instruction to generate responses grounded exclusively on the retrieved graph evidence without introducing unsupported external knowledge. This separation between symbolic retrieval and natural-language generation improves reasoning transparency while reducing hallucination, consistent with prior work on knowledge-grounded reasoning in RAG systems~\cite{fang2024, li2025agenticragdeepreasoning}.

\begin{algorithm}[t]
\footnotesize
\caption{KG-RAG Retrieval with Symbolic and Embedding Fallback}
\label{alg:kgrag-retrieval}
\begin{algorithmic}[1]
\Require User query $q$, ontology $\mathcal{O}$, knowledge graph $\mathcal{G} = (\mathcal{V}, \mathcal{E})$, Cypher LLM $\mathrm{LLM}_{\text{cypher}}$, encoder $f_{\text{enc}}$, top-$k$ parameter $k$
\Ensure Retrieved evidence set $R$ for downstream answer synthesis

\State \textbf{// Phase 1: Symbolic retrieval}
\State $\hat{C} \leftarrow \mathrm{LLM}_{\text{cypher}}(q, \mathcal{O})$ \Comment{Generate candidate Cypher queries}
\State Deduplicate $\hat{C}$

\ForAll{$\hat{c} \in \hat{C}$}
  \State $R \leftarrow \mathrm{EXEC}(\hat{c}, \mathcal{G})$
  \If{$R \neq \emptyset$}
    \State \textbf{return} $R$
  \EndIf
\EndFor

\State \textbf{// Phase 2: Embedding-based fallback}
\State $\mathbf{z}_q \leftarrow f_{\text{enc}}(q)$

\ForAll{$v_i \in \mathcal{V}$}
  \State $\mathbf{z}_i \leftarrow f_{\text{enc}}(v_i)$
  \State $s_i \leftarrow \mathrm{sim}(\mathbf{z}_q, \mathbf{z}_i)$ \Comment{e.g., cosine similarity}
\EndFor

\State $\mathcal{V}_k \leftarrow \mathrm{TopK}(\{(v_i, s_i)\}, k)$

\State \textbf{// Re-anchor symbolic retrieval}
\ForAll{$v \in \mathcal{V}_k$}
  \State $\hat{c}' \leftarrow \mathrm{LLM}_{\text{cypher}}(q, v, \mathcal{O})$
  \State $R' \leftarrow \mathrm{EXEC}(\hat{c}', \mathcal{G})$
  \If{$R' \neq \emptyset$}
    \State \textbf{return} $R'$
  \EndIf
\EndFor

\State \textbf{return} $\emptyset$
\end{algorithmic}
\end{algorithm}

\begin{algorithm}[!t]
\footnotesize
\caption{Prompt Segmentation and Cypher Generation}
\label{alg:prompt-cypher}
\begin{algorithmic}[1]
\Require User query $q$, ontology $\mathcal{O}$, maximum segments $S_{\max}$, Cypher LLM $\mathrm{LLM}_{\text{cypher}}$
\Ensure Set of Cypher queries $\hat{C}$
\If{length$(q)$ $\leq$ token threshold}
  \State $\mathcal{S} \leftarrow \{q\}$
\Else
  \State Segment $q$ into phrases $\mathcal{S} = \{s_1, \dots, s_S\}$ with $S \leq S_{\max}$
\EndIf
\State Construct prompt $P$ using:
\State \quad compact ontology schema from $\mathcal{O}$
\State \quad Cypher construction instructions
\State \quad segmented phrases wrapped with \texttt{CONTAINS} filters
\State $\hat{C} \leftarrow \mathrm{LLM}_{\text{cypher}}(P)$
\State Deduplicate $\hat{C}$ and discard syntactically invalid queries
\State \textbf{return} $\hat{C}$
\end{algorithmic}
\end{algorithm}

\subsection{Human--AI Collaborative Decision}

The final outcome of the proposed framework is an interactive human--AI system that enables users to query complex cyber-physical security knowledge through natural language. Users can submit high-level questions (e.g., related to vulnerabilities, attack paths, or system risks), which are processed through the hybrid retrieval and reasoning pipeline. The system integrates symbolic graph-based reasoning with embedding-supported retrieval to generate accurate and context-aware responses grounded in the knowledge graph.

The retrieved results are presented in an interpretable manner, allowing users to explore relationships between vulnerabilities, weaknesses, and attack techniques. This supports a range of decision-making tasks, including attack analysis, vulnerability identification, and mitigation planning. By combining structured graph reasoning with language model capabilities, the framework facilitates explainable and efficient Human--AI collaboration for cybersecurity analysis.

\section{Evaluation}
\label{sec:evaluation}

This section assesses GRICS in terms of benchmark performance, component contributions, adversarial robustness, runtime efficiency, and explainability. Four configurations are compared using identical benchmark inputs and inference settings: (i) Base KG-RAG, (ii) KG-RAG with fine-tuning (KG-RAG+FT), (iii) KG-RAG with embedding fallback (KG-RAG+EF), and (iv) Full GRICS. The analysis combines CTI-Benchmark tasks with multi-hop reasoning experiments to examine retrieval accuracy, robustness, computational performance, ontology consistency, and reasoning traceability.

\vspace{-1.3em}
\subsection{Experimental Setup}
\label{sec:fine}

The experimental setup comprises the construction of the KG-RAG question-answering dataset, the fine-tuning strategy for the Cypher-LLM, the training configuration, and the evaluation protocol for threat-centric retrieval and reasoning tasks.

\vspace{-0.3em}
\subsubsection{KG-RAG QA Dataset}
\label{subsec:dataset}

A supervised KG-RAG question-answering (QA) dataset was constructed to evaluate the effectiveness of the proposed framework in supporting threat-centric reasoning. The dataset was derived from 65 real-world CVEs selected from the BRIDG-ICS knowledge graph, each associated with complete CVE$\rightarrow$CWE$\rightarrow$CAPEC$\rightarrow$ATT\&CK mappings. From this set, multiple query instances were generated, resulting in a total of 450 samples aligned with the BRIDG-ICS ontology, enabling structured and ontology-aware retrieval and reasoning. The dataset captures diverse cybersecurity reasoning tasks, including: 
(i) inter-node relationships (e.g., CVE-CWE, CWE-CAPEC),
(ii) entity dependency reasoning,
(iii) vulnerability and attack technique explanations,
(iv) multi-hop graph traversal and attack path discovery,
(v) mitigation-oriented queries,
(vi) cross-domain IT–OT relationships, and
(vii) Vulnerability Propagation Risk (VPR) analysis.

Structural reasoning is encouraged while avoiding overfitting to specific identifiers through controlled paraphrasing and identifier variation. Each CVE instance is associated with multiple paraphrased query forms (e.g., ``Which CVEs are related to CWE-200'' and ``Find CVEs associated with CWE-200''), while CVE identifiers may be modified or substituted (e.g., \texttt{CVE-2024-30051}) to promote generalisation. Despite these variations, all question–answer pairs remain explicitly grounded in the BRIDG-ICS schema, ensuring consistency between natural language queries and graph-based representations.

\begin{table}[h!]
\centering
\small
\caption{Summary of KG-RAG QA fine-tuning dataset.}
\label{tab:finetune_summary}
\begin{tabular}{l c}
\toprule
\textbf{Category} & \textbf{Samples} \\
\midrule
CWE and CVE inter-node queries & 50 \\
Entity relations and traversal & 65 \\
Entity explanations by identifier & 150 \\
Entity explanations by name & 50 \\
Multi-hop graph traversal & 100 \\
Path dependency analysis & 35 \\
\midrule
\textbf{Total} & \textbf{450} \\
\bottomrule
\end{tabular}
\end{table}

\subsubsection{Model Training and Fine-Tuning}
\label{subsec:training}

The Cypher generation module is implemented using a pre-trained Llama-3.1-8B Text2Cypher model and adapted to the cybersecurity domain through parameter-efficient fine-tuning using the Unsloth framework, which provides memory-efficient optimisation for large language models. A total of 450 curated training samples constructed from the BRIDG-ICS knowledge graph were used to fine-tune the Cypher-LLM for Cypher generation and graph retrieval tasks. During inference, few-shot prompting is employed to improve syntactic correctness and semantic consistency by constraining generated Cypher queries to valid node labels, relationship types, and property names defined in the BRIDG-ICS ontology. Specifically, the ontology schema is explicitly provided as part of the prompt context, allowing the model to generate ontology-compliant Cypher queries by adapting to the supplied graph schema during inference rather than requiring the complete ontology structure to be encoded in the model parameters.

The Cypher-LLM is fine-tuned to enhance the translation of natural-language cybersecurity queries into executable Cypher statements and to align the model with domain-specific ontology structures and query patterns. In contrast, the Answer-LLM remains prompt-based to preserve flexibility and avoid over-specialisation to a fixed response format. Formally, the Cypher generation process is defined by Equation~(\ref{eq:mapping}):

\begin{equation}
\hat{c}=\mathcal{F}_{\theta}(q,\mathcal{O})
\label{eq:mapping}
\end{equation}

where $q$ denotes the input natural-language query, $\mathcal{O}$ represents the BRIDG-ICS ontology schema, $\mathcal{F}_{\theta}$ denotes the fine-tuned Cypher generation model parameterised by $\theta$, and $\hat{c}$ is the generated executable Cypher query.

This objective encourages the generation of syntactically valid and semantically consistent Cypher queries while reducing reliance on extensive prompt engineering. Fine-tuning further improves robustness under incomplete graph structures by enabling alternative graph traversal strategies when direct relationships are unavailable. Pathfinding and traversal operations are implemented using graph data science techniques, enabling efficient multi-hop reasoning and attack-path exploration within the cybersecurity knowledge graph. The model converged with a final training loss of \textbf{0.0052}, indicating stable optimisation and effective alignment between natural-language instructions and symbolic query generation.

The complete prompt templates, fine-tuning implementation, and data generation pipeline are publicly available in the project repository.\footnote{\url{https://github.com/ahmadspm/Resellient-Industry-5.0--kG-Digital-Twins}}

\begin{table}[h!]
\centering
\caption{Training configuration for Cypher-LLM fine-tuning.}
\label{tab:train_config}
\begin{tabular}{ll}
\hline
\textbf{Parameter} & \textbf{Value} \\
\hline
per device train batch size & 1 \\
gradient accumulation steps & 4 \\
num train epochs & 5 \\
learning rate & $2 \times 10^{-5}$ \\
bf16 & True \\
logging steps & 40 \\
save strategy & epoch \\
remove unused columns & False \\
gradient checkpointing & False \\
\hline
\end{tabular}
\end{table}

During inference, ontology-guided few-shot prompting is used to improve the syntactic validity and semantic consistency of generated Cypher queries. The prompt structure, validation process, and evidence-grounded response synthesis are described in Section~\ref{subsec:prompt_engineering}.

\vspace{-0.5em}
\subsubsection{Prompt Engineering and Inference Constraints}
\label{subsec:prompt_engineering}

GRICS uses separate prompts for the Cypher-LLM and Answer-LLM. The Cypher-LLM translates natural-language requests into executable Cypher queries, whereas the Answer-LLM synthesises responses from retrieved graph evidence.

The Cypher prompt includes task instructions, the relevant BRIDG-ICS schema, few-shot examples, and the analyst query, while restricting generated queries to the supplied node labels, relationships, and properties and producing up to three candidates per request. These candidates are evaluated sequentially for syntax, ontology conformity, and executability over the LPG, with the first valid query that retrieves relevant evidence being selected. When symbolic retrieval fails, embedding retrieval identifies semantically related graph anchors, which are supplied to the Cypher-LLM for query regeneration; the regenerated query must still pass validation and symbolic execution.

The Answer-LLM receives the original query and the retrieved graph evidence. Its prompt restricts response generation to the supplied entities, relationships, properties, and graph paths, ensuring that the final output remains grounded and traceable.

\subsubsection{Training Configuration}
\label{subsec:config}

Fine-tuning experiments were conducted on a workstation equipped with an NVIDIA RTX 6000 Ada Generation GPU (48\,GB VRAM). The model was trained using the Unsloth optimisation framework with Low-Rank Adaptation (LoRA)-based adapters. The training configuration includes a sequence length of 2048 tokens and an effective batch size of 4 achieved via gradient accumulation. The key hyperparameters are summarised in Table~\ref{tab:train_config}.



\subsubsection{Benchmark Datasets and Evaluation Protocol}
\label{subsec:benchmark_protocol}

The proposed framework was evaluated using the publicly available CTI-Benchmark (CTIBench)~\cite{ctibench}, which provides standardised cybersecurity reasoning tasks for evaluating large language models. Three benchmark tasks were selected in this study: CTI-RCM 2024, CTI-RCM 2021, and CTI-ATE. The CTI-RCM benchmarks evaluated the classification of CVE descriptions into their corresponding CWE categories using two temporal splits, while CTI-ATE evaluated the identification of MITRE ATT\&CK techniques associated with software or malware entities.

Since CTI-Benchmark is originally designed for prompt-based question answering, the benchmark prompts were reformulated into natural-language questions compatible with the proposed KG-RAG framework. For example, a vulnerability classification prompt was reformulated as: ``What is the CWE associated with the CVE described as `...' ?'' This reformulation enables the Cypher-LLM to translate natural-language questions into executable Cypher queries while preserving the original benchmark objectives.
Benchmark evaluation focuses on the Cypher generation and symbolic graph retrieval stages of GRICS. Specifically, the benchmark measures whether the Cypher-LLM correctly maps the natural-language query to the corresponding cybersecurity entity through graph retrieval. This evaluation protocol is appropriate because the benchmark ground truth consists of a single structured entity (e.g., a CWE identifier or MITRE ATT\&CK technique), which is determined before response synthesis. The subsequent Answer-LLM is responsible only for transforming the retrieved graph evidence into a human-readable explanation and does not alter the retrieved entity. Consequently, benchmark performance reflects the correctness of the graph-grounded retrieval and reasoning process rather than the natural-language generation stage.

The Base KG-RAG configuration employs prompt-based Cypher generation together with symbolic retrieval over the BRIDG-ICS knowledge graph. The KG-RAG+FT configuration additionally enables Cypher fine-tuning, KG-RAG+EF enables the embedding-based fallback mechanism, and Full GRICS combines both components while maintaining the same symbolic retrieval pipeline.

Performance is evaluated using Accuracy, Precision, Recall, and F1-score for the CTI-Benchmark tasks. Adversarial robustness is evaluated using Attack Success Rate (ASR) and Tokens per Query (TPQ), while explainability is assessed using Hallucination Rate (HR), Query Violation Rate (QVR), and Schema Consistency Rate (SCR). Identical prompts, inference settings, and evaluation metrics are maintained across all configurations to ensure fair and reproducible comparison.

\subsection{Benchmark Performance Comparison}
\label{subsec:benchmark_performance}

This section evaluates the performance of GRICS on three tasks from the CTI-Benchmark: CTI-RCM 2024, CTI-RCM 2021, and CTI-ATE. The CTI-RCM tasks evaluate the ability of the framework to associate vulnerability descriptions with their corresponding Common Weakness Enumeration (CWE) categories, whereas CTI-ATE evaluates the identification of MITRE ATT\&CK techniques associated with software or malware entities. Collectively, these tasks assess ontology-aware entity retrieval, structured cybersecurity knowledge reasoning, and graph-grounded threat analysis.

Table~\ref{tab:unified_performance} presents the performance of the evaluated GRICS configurations across the three benchmark tasks. Full GRICS achieves the highest performance on all three datasets, obtaining an accuracy of 87.60\% and an F1 score of 0.930 on CTI-RCM 2024, an accuracy of 90.40\% and an F1 score of 0.949 on CTI-RCM 2021, and an accuracy of 77.15\% with an F1 score of 0.871 on CTI-ATE.
The results show that GRICS performs strongly on the two vulnerability-to-weakness classification tasks, while the lower CTI-ATE performance reflects the greater relational complexity of ATT\&CK technique identification across software, malware, attack patterns, vulnerabilities, and techniques. Nevertheless, Full GRICS remains effective for deeper graph traversal and heterogeneous reasoning. Precision is 100\% across all configurations, indicating that valid predictions retrieve the correct entity; therefore, differences in accuracy, recall, and F1 score mainly reflect each configuration's ability to resolve queries and retrieve sufficient graph-grounded evidence.

\begin{table}[t]
\centering
\scriptsize
\setlength{\tabcolsep}{3pt}
\renewcommand{\arraystretch}{0.95}
\caption{Component-wise evaluation of GRICS on the CTI-Benchmark.}
\label{tab:unified_performance}
\begin{tabular}{lcccc}
\toprule
\multicolumn{5}{c}{\textbf{CTI-RCM 2024}} \\
\midrule
\textbf{Model} & \textbf{Acc.} & \textbf{Prec.} &
\textbf{Rec.} & \textbf{F1} \\
\midrule
Base KG-RAG  & 62.80 & 100 & 62.80 & 0.771 \\
KG-RAG + FT  & 67.20 & 100 & 67.20 & 0.800 \\
KG-RAG + EF  & 85.70 & 100 & 85.70 & 0.923 \\
Full GRICS   & \textbf{87.60} & 100 & \textbf{87.60} &
\textbf{0.930} \\
\midrule
\multicolumn{5}{c}{\textbf{CTI-RCM 2021}} \\
\midrule
Base KG-RAG  & 66.50 & 100 & 66.50 & 0.798 \\
KG-RAG + FT  & 78.60 & 100 & 78.60 & 0.880 \\
KG-RAG + EF  & 88.20 & 100 & 88.20 & 0.937 \\
Full GRICS   & \textbf{90.40} & 100 & \textbf{90.40} &
\textbf{0.949} \\
\midrule
\multicolumn{5}{c}{\textbf{CTI-ATE}} \\
\midrule
Base KG-RAG  & 18.33 & 100 & 18.33 & 0.309 \\
KG-RAG + FT  & 64.72 & 100 & 64.72 & 0.786 \\
KG-RAG + EF  & 53.33 & 100 & 53.33 & 0.695 \\
Full GRICS   & \textbf{77.15} & 100 & \textbf{77.15} &
\textbf{0.871} \\
\bottomrule
\end{tabular}
\end{table}

The benchmark results demonstrate that Full GRICS performs effectively across vulnerability classification and ATT\&CK technique identification. The following ablation study further examines the contributions of graph grounding and the individual architectural components.

\paragraph{\textbf{Statistical Significance}.}

Pairwise statistical comparisons were performed using McNemar's test at a significance level of $\alpha=0.05$ for each of the three benchmark tasks: CTI-RCM 2024, CTI-RCM 2021, and CTI-ATE. On each benchmark, the published Base LLM was compared separately with Base KG-RAG, KG-RAG+FT, KG-RAG+EF, and Full GRICS. The results showed that Base KG-RAG significantly outperformed the Base LLM across CTI-RCM 2024, CTI-RCM 2021, and CTI-ATE ($p<0.05$), demonstrating the benefit of symbolic knowledge-graph retrieval. KG-RAG+FT also achieved statistically significant improvements over the Base LLM on all three benchmarks ($p<0.05$), indicating the contribution of domain-specific Cypher fine-tuning. Similarly, KG-RAG+EF significantly outperformed the Base LLM across the three tasks ($p<0.05$), highlighting the effectiveness of embedding-assisted query recovery. Full GRICS achieved the strongest performance and significantly outperformed the Base LLM on CTI-RCM 2024, CTI-RCM 2021, and CTI-ATE ($p<0.05$), demonstrating the combined benefit of symbolic graph retrieval, Cypher fine-tuning, and  CC BY: Creative Commons Attribution    CC BY-SA: Creative Commons Attribution-ShareAlike    CC BY-NC-SA: C embedding fallback.

\subsubsection{Ablation Study}
\label{subsubsec:ablation}

The ablation study analyses the contribution of graph-grounded symbolic retrieval, Cypher fine-tuning, and embedding-based fallback. Since graph grounding in GRICS is realised through symbolic Cypher execution over the BRIDG-ICS knowledge graph, it cannot be removed while preserving an executable KG-RAG configuration. Therefore, the contribution of graph grounding is assessed by comparing the published CTI-Benchmark LLM baseline, which operates without knowledge graph grounding, with Base KG-RAG.
It uses prompt-based Cypher generation followed by symbolic execution over the BRIDG-ICS knowledge graph. The difference between the published LLM baseline and Base KG-RAG therefore represents the contribution of graph-grounded symbolic retrieval. The remaining configurations progressively introduce Cypher fine-tuning (KG-RAG + FT), embedding-based fallback (KG-RAG + EF), and their combination in Full GRICS.

\begin{figure*}[t]
    \centering

    \begin{subfigure}[t]{0.32\textwidth}
        \centering
        \includegraphics[width=\linewidth]
        {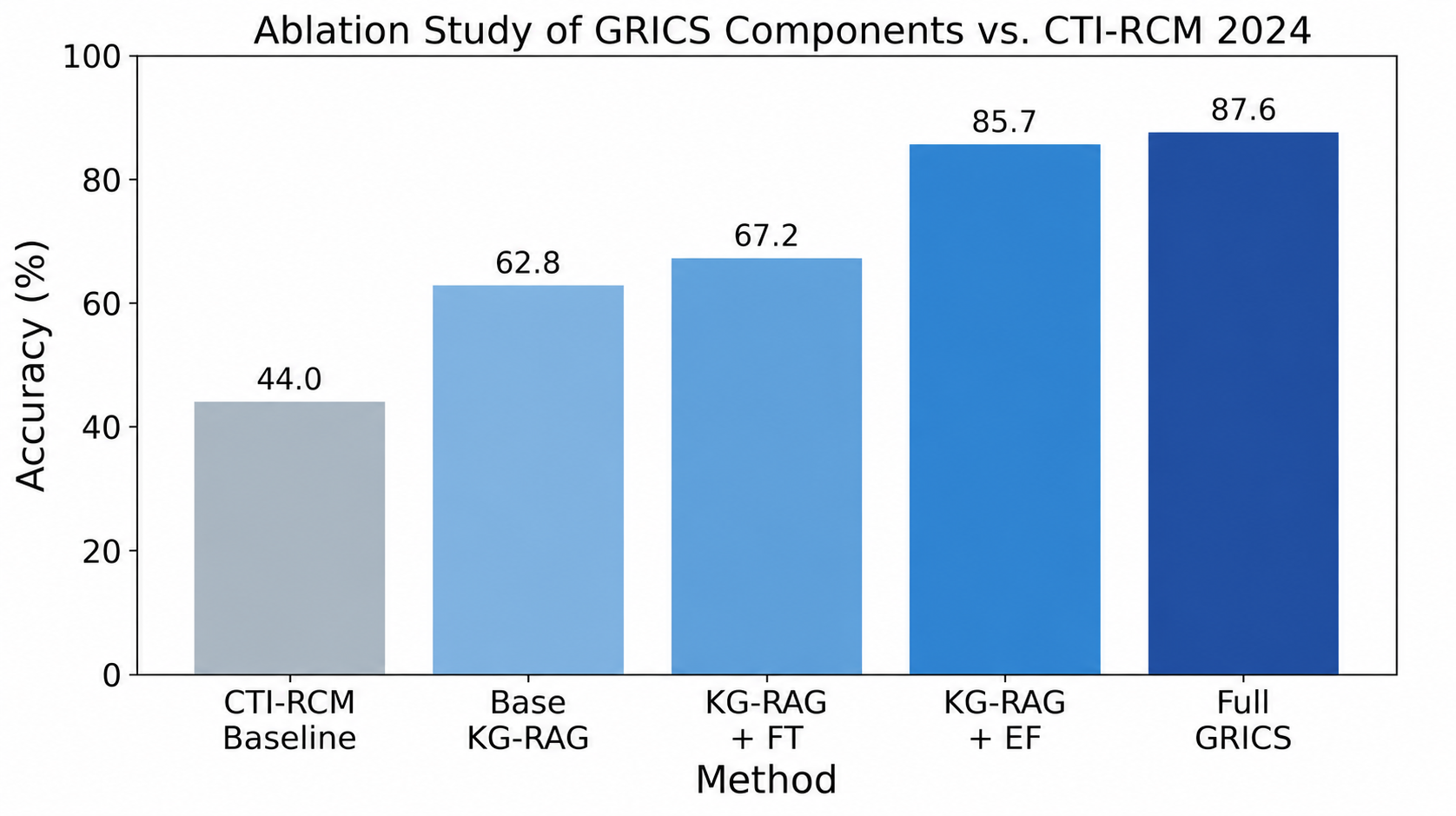}
        \caption{CTI-RCM 2024}
        \label{fig:acc_cti_rcm2024}
    \end{subfigure}%
    \hfill
    \begin{subfigure}[t]{0.32\textwidth}
        \centering
        \includegraphics[width=\linewidth]
        {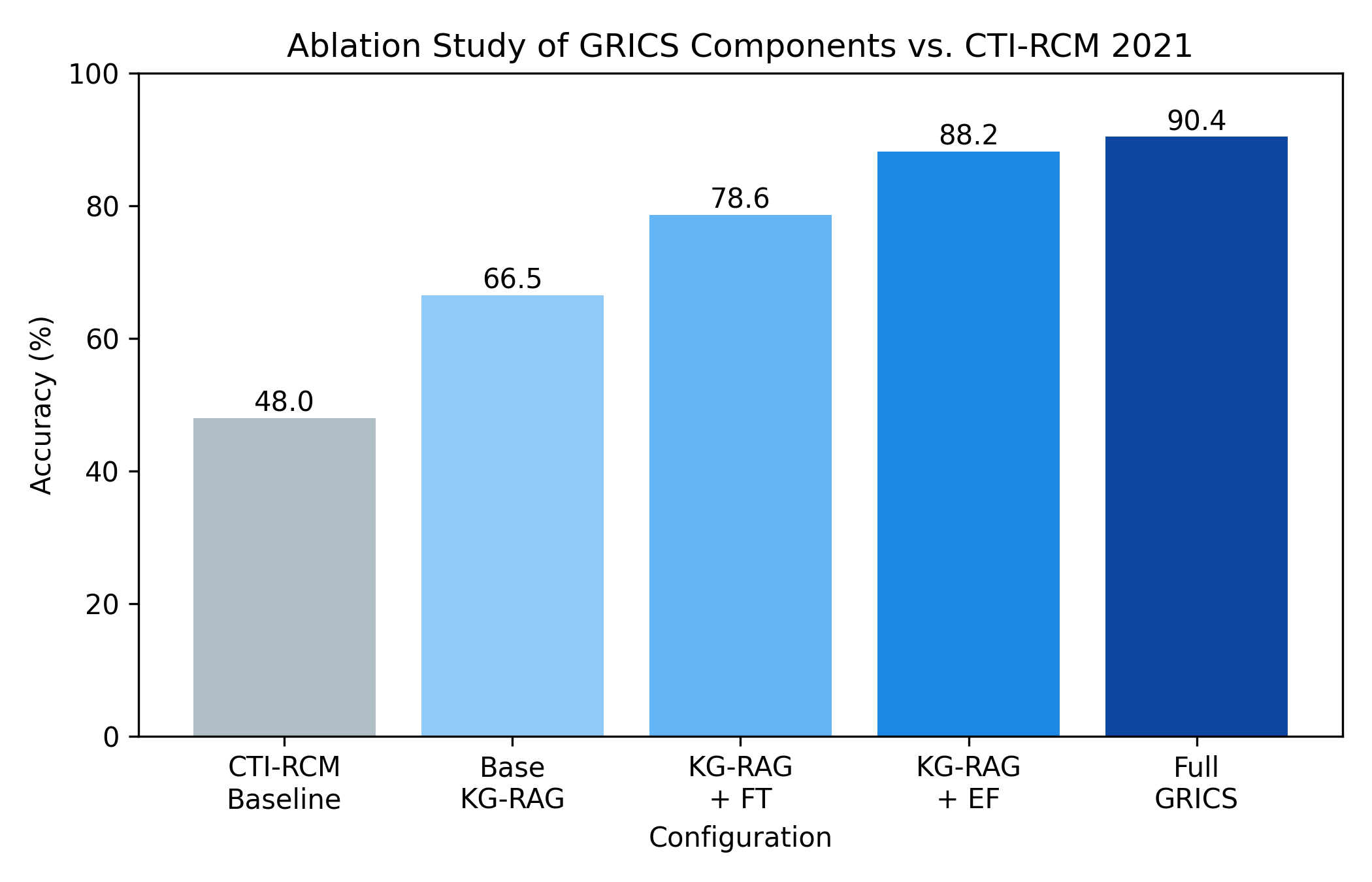}
        \caption{CTI-RCM 2021}
        \label{fig:acc_cti_rcm2021}
    \end{subfigure}%
    \hfill
    \begin{subfigure}[t]{0.32\textwidth}
        \centering
        \includegraphics[width=\linewidth]
        {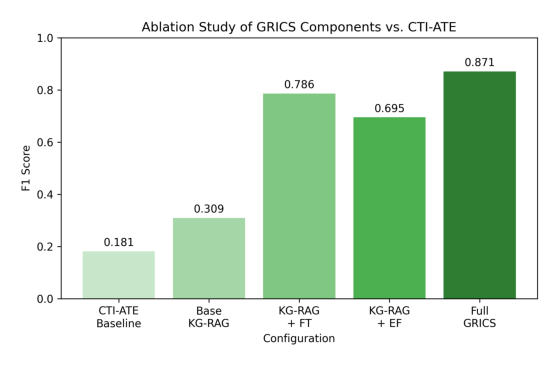}
        \caption{CTI-ATE}
        \label{fig:acc_cti_ate}
    \end{subfigure}

    \caption{Component-wise ablation of GRICS on the CTI-Benchmark
    datasets. The published CTI-Benchmark LLM baseline represents
    prediction without knowledge graph grounding. Base KG-RAG employs
    prompt-based Cypher generation and symbolic graph retrieval, +FT
    enables Cypher fine-tuning, +EF enables the embedding-based fallback
    mechanism, and Full GRICS combines both enhancements.}
    \label{fig:benchmark_comparison}
\end{figure*}

Figure~\ref{fig:benchmark_comparison} illustrates the contribution of each component across the three benchmark tasks. Compared with the published LLM baseline, Base KG-RAG demonstrates the benefit of graph-grounded retrieval, while Cypher fine-tuning further improves query validity and semantic alignment, particularly on the more relationally complex CTI-ATE task.

The embedding-based fallback mechanism improves retrieval coverage when the initial Cypher query is invalid, overly restrictive, or produces no relevant result. Importantly, the fallback mechanism does not independently generate the final answer or replace symbolic graph reasoning. Instead, it identifies semantically related graph anchors and supplies them to the Cypher generation module, which regenerates an ontology-compliant query before symbolic execution resumes.

Full GRICS achieves the strongest performance across all three benchmark tasks, demonstrating that Cypher fine-tuning and embedding-based fallback address complementary limitations. Fine-tuning improves the quality of the initial symbolic query, whereas embedding-assisted rec  CC BY: Creative Commons Attribution    CC BY-SA: Creative Commons Attribution-ShareAlike    CC BY-NC-SA: Covery improves robustness when the initial retrieval attempt is unsuccessful. These ablation results show that GRICS’s performance gains stem from three complementary capabilities: explicit knowledge graph grounding, task-specific Cypher generation, and embedding-assisted query recovery. Comparison with the published CTI-Benchmark LLM baseline quantifies the impact of graph-grounded symbolic retrieval, while internal configuration comparisons isolate the added benefits of Cypher fine-tuning and the embedding-based fallback.

\paragraph{\textbf{Explainability Analysis.}}

\begin{figure}[t]
    \centering

    \begin{subfigure}[t]{0.48\columnwidth}
        \centering
        \includegraphics[width=\linewidth]
        {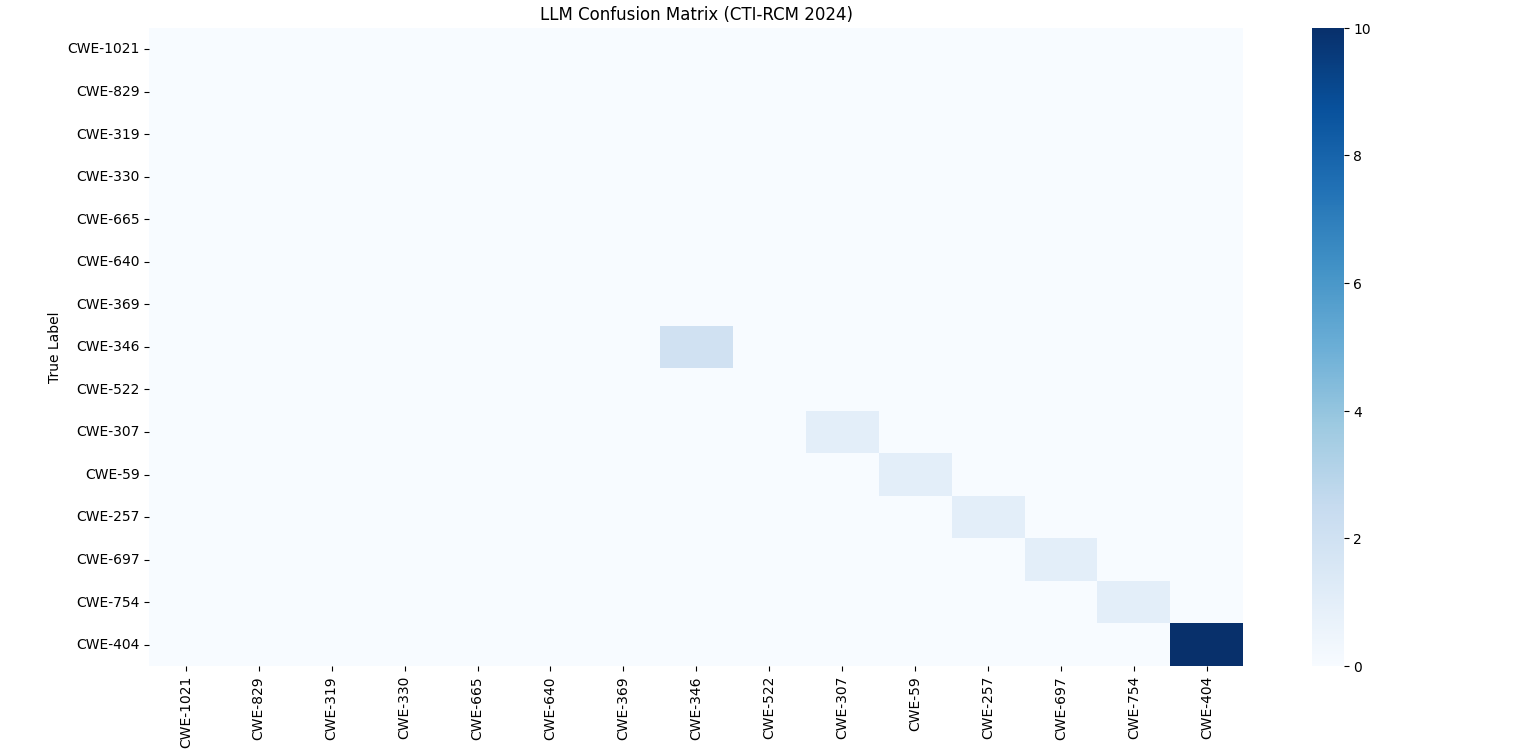}
        \caption{CTI-RCM 2024 Base KG-RAG}
    \end{subfigure}
    \hfill
    \begin{subfigure}[t]{0.48\columnwidth}
        \centering
        \includegraphics[width=\linewidth]
        {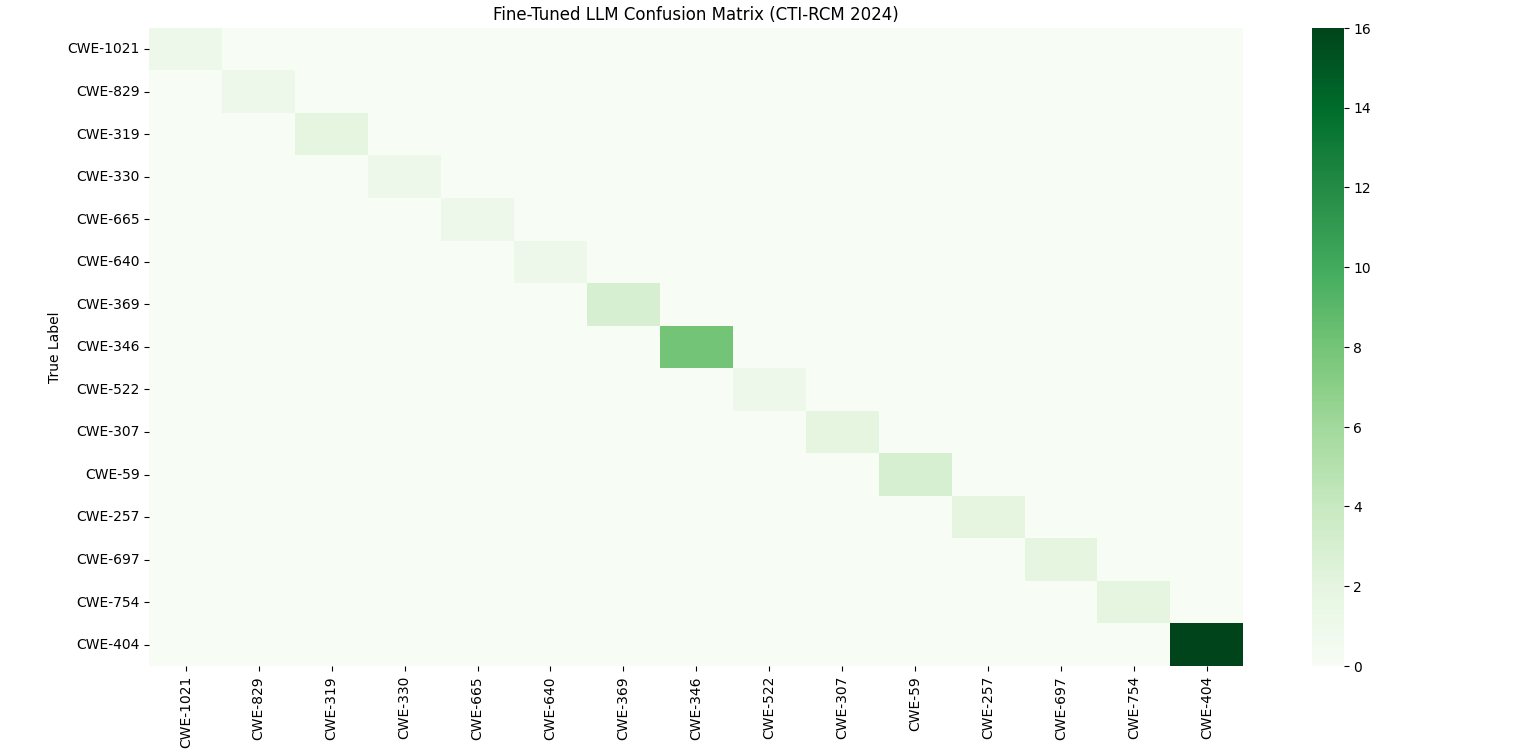}
        \caption{CTI-RCM 2024 KG-RAG + FT}
    \end{subfigure}

    \vspace{0.3em}

    \begin{subfigure}[t]{0.48\columnwidth}
        \centering
        \includegraphics[width=\linewidth]
        {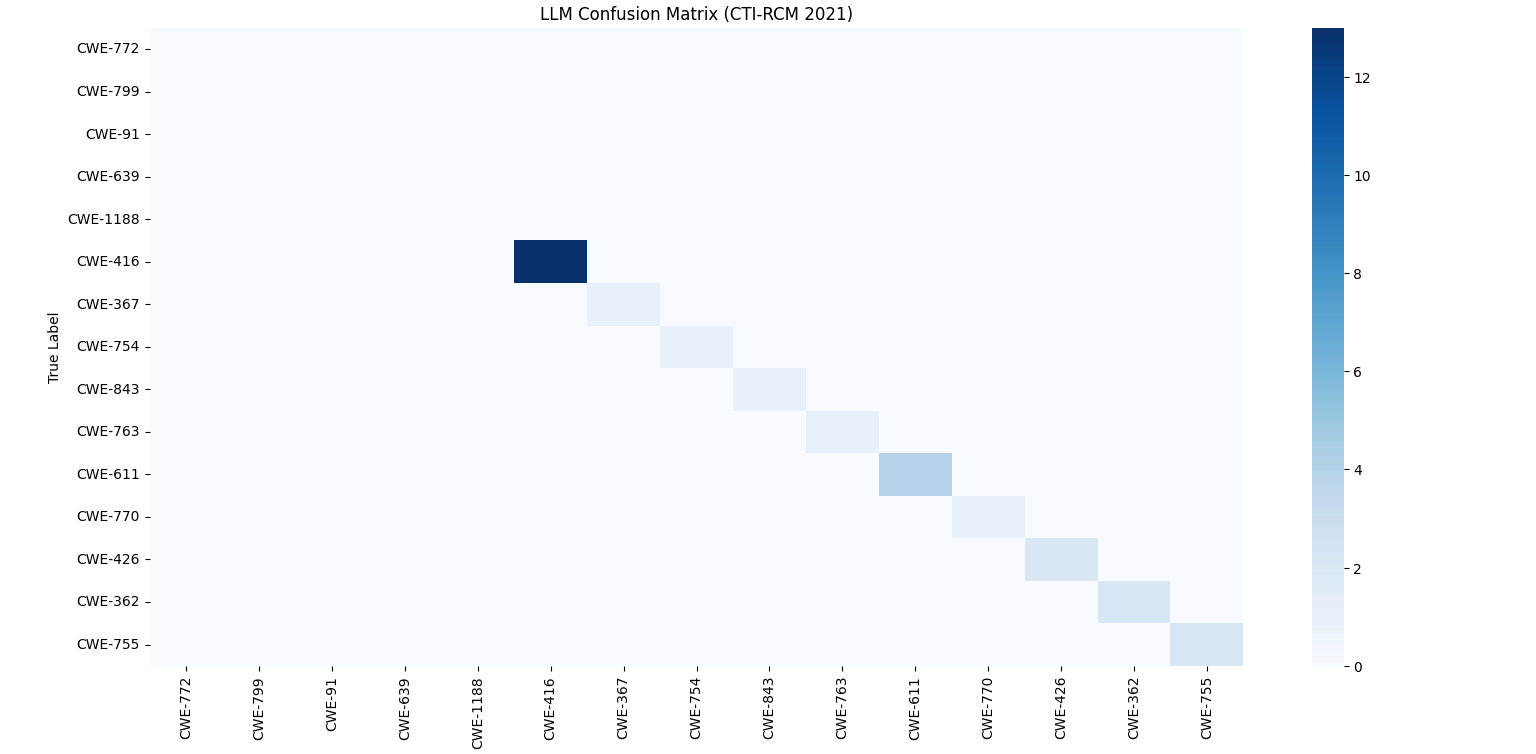}
        \caption{CTI-RCM 2021 Base KG-RAG}
    \end{subfigure}
    \hfill
    \begin{subfigure}[t]{0.48\columnwidth}
        \centering
        \includegraphics[width=\linewidth]
        {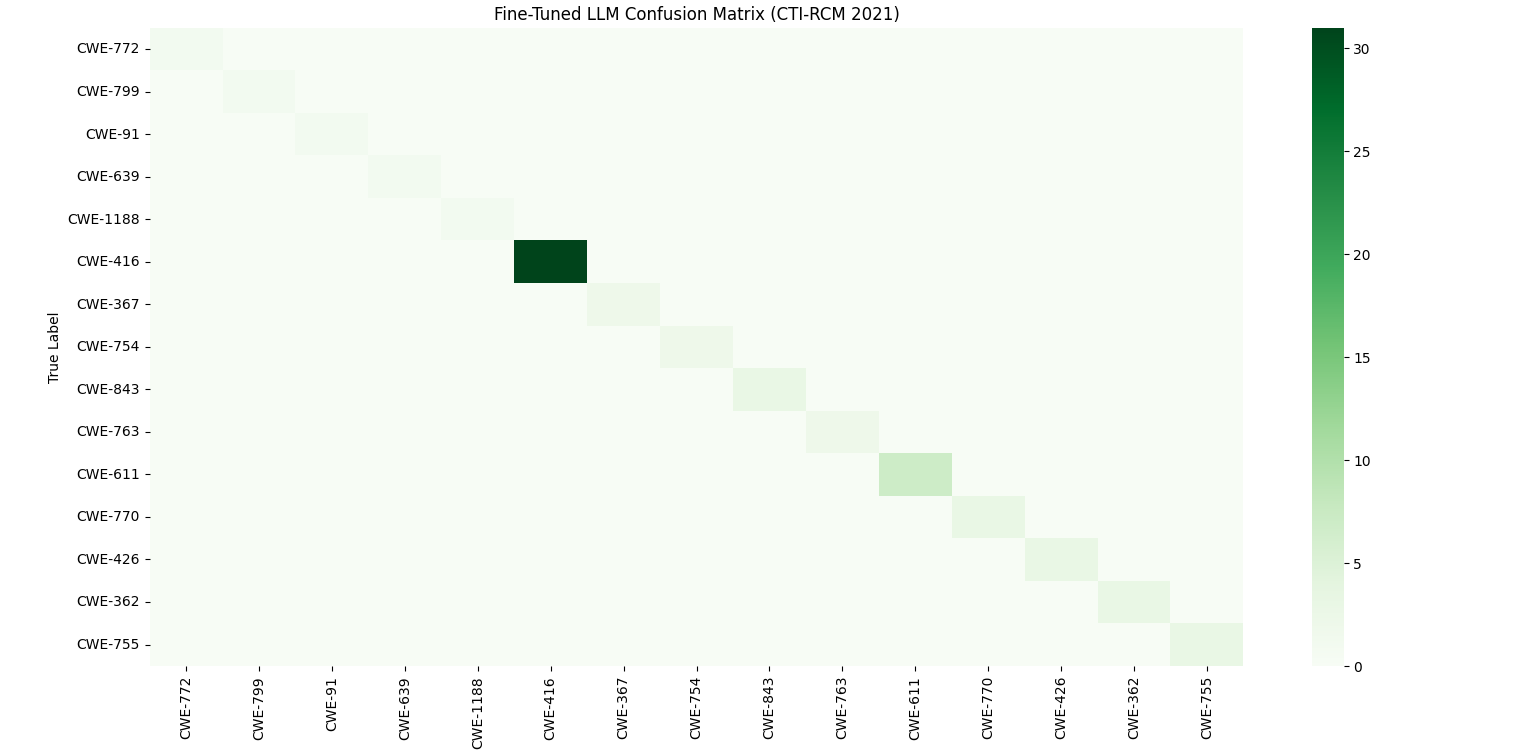}
        \caption{CTI-RCM 2021 KG-RAG + FT}
    \end{subfigure}

    \caption{Heatmap comparison of predictive accuracy across the top 15 CWE classes for CTI-RCM 2024 and CTI-RCM 2021.}
    \label{fig:heatmaps1}
\end{figure}

\begin{figure*}[t]
    \centering

    \begin{subfigure}[t]{0.28\textwidth}
        \centering
        \includegraphics[width=\linewidth]
        {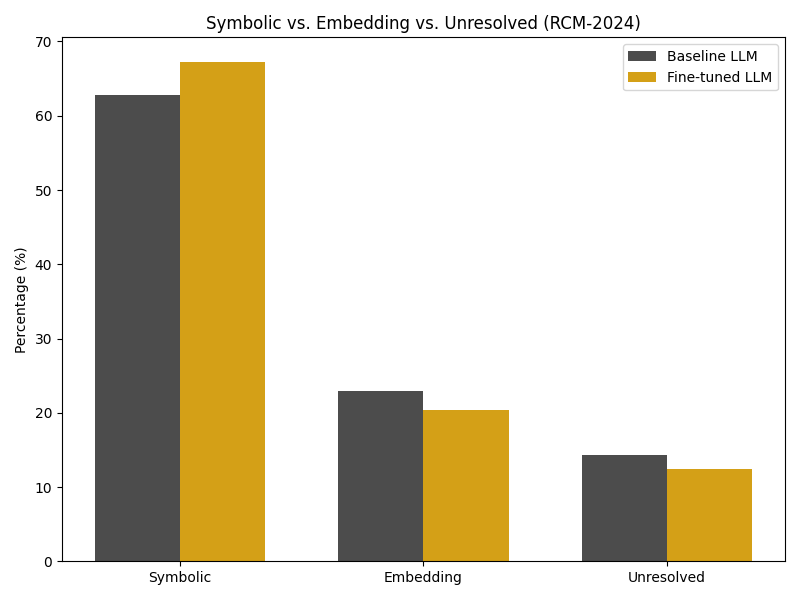}
        \caption{CTI-RCM 2024}
    \end{subfigure}
    \hfill
    \begin{subfigure}[t]{0.28\textwidth}
        \centering
        \includegraphics[width=\linewidth]
        {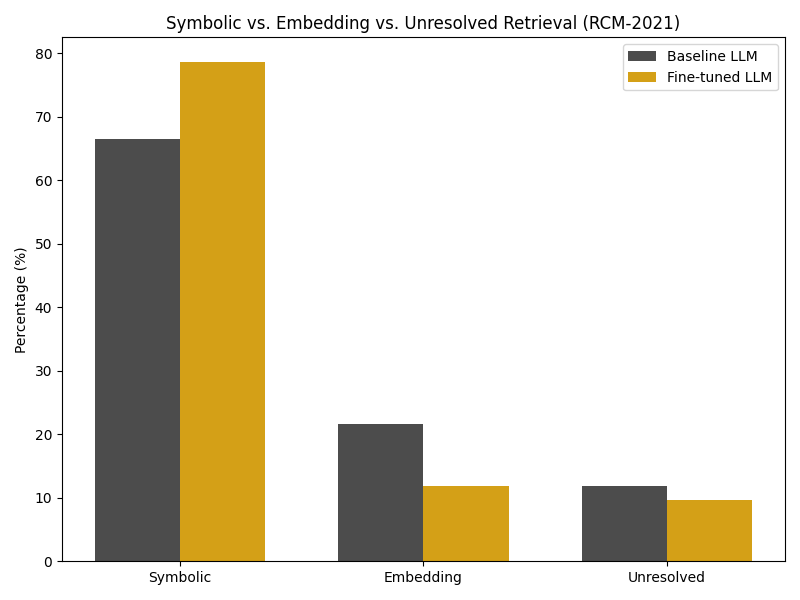}
        \caption{CTI-RCM 2021}
    \end{subfigure}
    \hfill
    \begin{subfigure}[t]{0.28\textwidth}
        \centering
        \includegraphics[width=\linewidth]
        {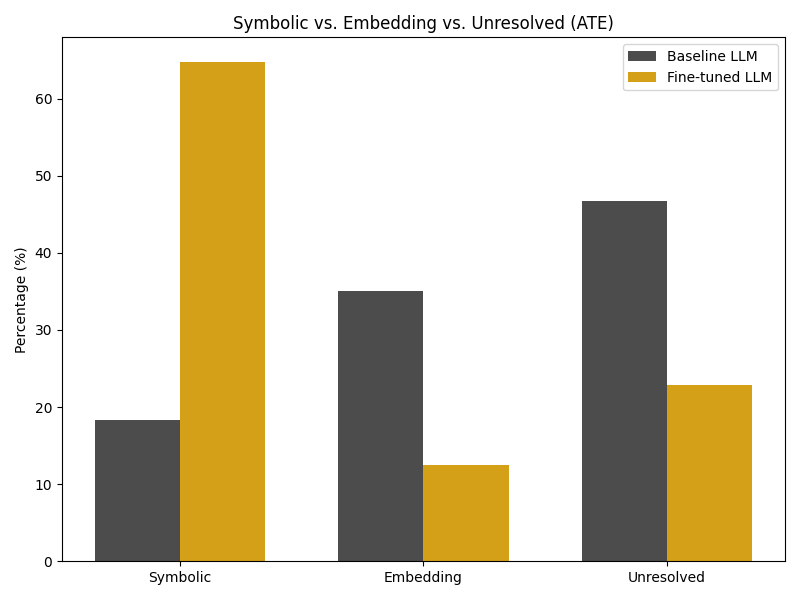}
        \caption{CTI-ATE}
    \end{subfigure}

    \caption{Retrieval outcome distributions across the CTI-RCM 2024, CTI-RCM 2021, and CTI-ATE benchmarks, showing symbolic-only retrievals, embedding-assisted fallback retrievals, and unresolved queries for the evaluated GRICS configurations.}
    \label{fig:symbolic_percentage}
\end{figure*}

The explainability of GRICS is examined from two complementary perspectives: semantic coverage and retrieval transparency. While Figure~\ref{fig:benchmark_comparison} demonstrates the performance contribution of the individual architectural components, Figures~\ref{fig:heatmaps1} and~\ref{fig:symbolic_percentage} provide further insight into how these components influence the reasoning behaviour of the framework.
Figure~\ref{fig:heatmaps1} presents CWE-level prediction coverage for the CTI-RCM 2024 and CTI-RCM 2021 benchmarks. The Base KG-RAG configuration exhibits limited coverage and inconsistent predictions across several CWE classes. In contrast, KG-RAG + FT produces broader semantic coverage and more consistent predictions, particularly for less frequent and semantically complex CWE categories. This improvement indicates that fine-tuning strengthens the alignment between natural-language vulnerability descriptions, generated Cypher queries, and the structured cybersecurity concepts represented in the BRIDG-ICS knowledge graph.
\newline 
\textbf{\textit{Retrieval-level explainability}} is further analysed through the retrieval outcome distributions shown in Figure~\ref{fig:symbolic_percentage}. The distributions distinguish queries resolved through direct symbolic retrieval, queries recovered through the embedding-based fallback mechanism, and unresolved queries.

The fine-tuned configurations resolve a larger proportion of queries through direct symbolic reasoning and produce fewer unresolved outcomes. This behaviour improves reasoning traceability because more predictions can be associated with an explicit Cypher query and a corresponding graph traversal path. The embedding-based fallback mechanism provides an additional recovery path when the initial symbolic query is unsuccessful, but the final result remains grounded in symbolic graph execution because the retrieved semantic anchor is used to regenerate an ontology-compliant Cypher query. These results show that the architectural enhancements improve more than predictive performance. Cypher fine-tuning increases semantic coverage and the share of queries resolved by direct graph traversal, while embedding-assisted recovery reduces unresolved cases without replacing symbolic reasoning. GRICS thus offers an interpretable reasoning pipeline where retrieved evidence is traced to explicit entities and relationships in the BRIDG-ICS knowledge graph.

\paragraph{\textbf{\textit{Semantic Stability of GE Embeddings}}.}

The semantic stability of the graph embeddings used by the fallback mechanism is evaluated when the initial symbolic retrieval attempt is unsuccessful. Semantic stability refers to the consistency of similarity-based candidate rankings and the preservation of meaningful neighbourhood structures in the embedding space. All graph-node and query embeddings were generated using the CTI-RCM model.
Three complementary analyses are conducted: similarity-score decay across retrieval ranks, similarity margins between the two highest-ranked candidates, and a two-dimensional principal component analysis (PCA) projection. These analyses respectively examine rank-based relevance decay, candidate separation, and local neighbourhood structure.

\begin{figure*}[t]
    \centering

    \begin{subfigure}[t]{0.33\textwidth}
        \centering
        \includegraphics[height=3cm]
        {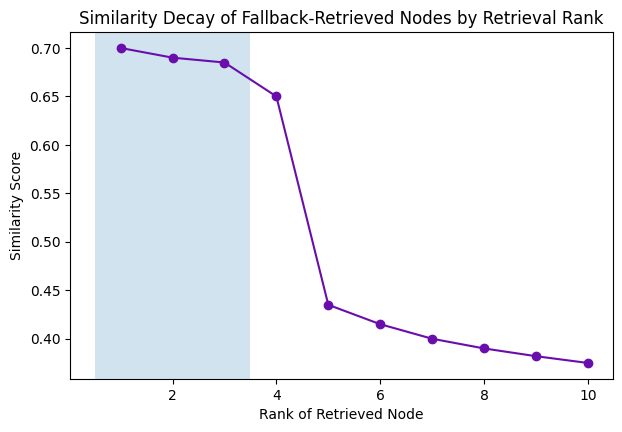}
        \caption{Similarity decay across retrieval rank.}
        \label{fig:similarity_decay}
    \end{subfigure}
    \hfill
    \begin{subfigure}[t]{0.33\textwidth}
        \centering
        \includegraphics[height=3cm]
        {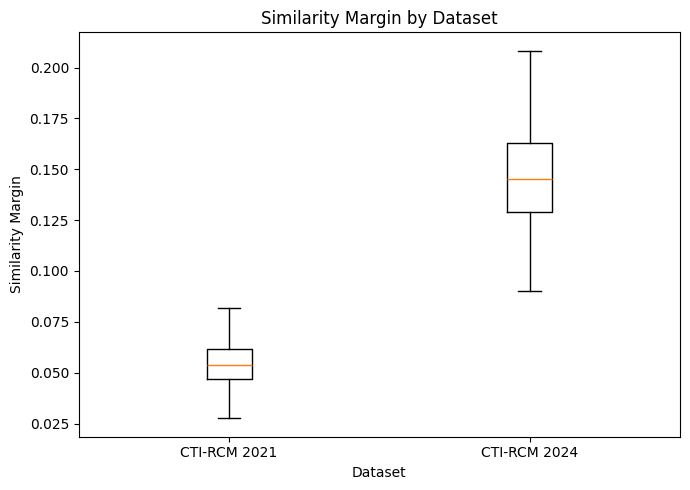}
        \caption{Similarity margin in embedding-based fallback retrieval.}
        \label{fig:hybrid_similarity}
    \end{subfigure}
    \hfill
    \begin{subfigure}[t]{0.33\textwidth}
        \centering
        \includegraphics[height=3cm]
        {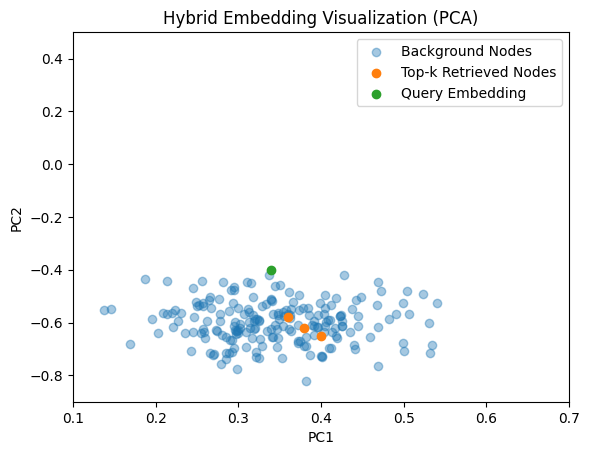}
        \caption{PCA projection of GE embeddings.}
        \label{fig:pca_ge}
    \end{subfigure}

    \caption{Semantic stability analysis of the graph embeddings used by the GRICS fallback retrieval mechanism.}
    \label{fig:ge_semantic_stability}
\end{figure*}

Figure~\ref{fig:similarity_decay} shows a clear decline in cosine similarity across retrieval ranks, based on 20 queries from each of the CTI-RCM 2021 and CTI-RCM 2024 datasets, indicating that the embedding model consistently prioritises a small set of semantically relevant graph entities. Figure~\ref{fig:hybrid_similarity} further evaluates candidate separation using the margin $\Delta = s_{\mathrm{top1}} - s_{\mathrm{top2}}$, where smaller margins for CTI-RCM 2021 indicate greater ambiguity between the highest-ranked candidates, while larger margins for CTI-RCM 2024 suggest stronger separation and more reliable graph-anchor selection.

The PCA projection in Figure~\ref{fig:pca_ge} provides a qualitative representation of the embedding neighbourhood for CTI-RCM 2024. The query embedding appears near its highest-ranked graph nodes and forms a coherent local cluster with semantically relevant entities. Although PCA does not preserve exact distances from the original high-dimensional space, the observed clustering provides qualitative evidence that relevant graph nodes occupy a similar semantic neighbourhood to the query.
The apparent proximity of some background nodes in the PCA projection does not contradict the retrieval rankings because candidate selection is performed using cosine similarity in the original embedding space rather than Euclidean distance in the two-dimensional projection.

Based on the observed similarity decay and candidate-margin separation, fallback retrieval in GRICS is restricted to the three highest-ranked graph nodes. The results indicate diminishing semantic relevance beyond these candidates. Limiting retrieval to the top three nodes therefore reduces the introduction of noisy graph anchors while preserving the candidates most likely to support successful ontology-aware Cypher regeneration.

\subsection{Robustness Against Adversarial Attacks}

Evaluating the robustness of KG-RAG systems against adversarial attacks remains challenging due to the absence of standardised benchmarks and the diversity of knowledge graph structures, ontology designs, and reasoning tasks adopted across existing studies. To assess the robustness of GRICS, we follow the evaluation methodology of \cite{liang2025graphrag} and adopt \textit{Attack Success Rate} (ASR) as the primary evaluation metric.For untargeted attacks, Attack Success Rate (ASR) is computed using Equation~(\ref{eq:asr}):

\begin{equation}
\label{eq:asr}
\text{ASR}=\frac{\sum_{(x,y)\in X}\mathbb{1}_{\hat{y}\neq y}}{|X|}.
\end{equation}

where $X$ denotes the set of evaluation queries, $y$ is the ground-truth answer, $\hat{y}$ is the model prediction under adversarial perturbation, and $\mathbb{1}(\cdot)$ is the indicator function that returns 1 when the attack successfully changes the prediction and 0 otherwise. A lower ASR indicates greater robustness, as fewer adversarial attacks successfully alter the model's prediction.

Tokens per Query (TPQ) is computed according to Equation~(\ref{eq:tpq}):

\begin{equation}
\text{TPQ}=\frac{\#\text{tokens in }D_{\text{poison}}}{|X|}.
\label{eq:tpq}
\end{equation}

where $D_{\text{poison}}$ represents the injected adversarial content and $|X|$ is the number of evaluated queries. TPQ measures the average number of adversarial tokens injected per query. Larger TPQ values indicate that an attacker must inject more adversarial content to influence the retrieval process, whereas lower TPQ values indicate that successful attacks require fewer injected tokens.

Following the experimental methodology of \cite{hotpotqa}, a total of 120 representative multi-hop reasoning queries were evaluated over the BRIDG-ICS ontology. To improve comparability with PoisonRAG and GRAPPOISON, the robustness evaluation was restricted to ATT\&CK-oriented reasoning tasks involving malware, ATT\&CK techniques, and mitigation relationships, reflecting the common evaluation scope adopted in prior graph-poisoning studies. Each query was subjected to adversarial perturbations following the prompt-injection strategy described in \cite{liang2025graphrag}.

The perturbations were grouped into lexical and semantic attack categories to assess whether different forms of input manipulation affect GRICS differently. Lexical attacks modify the surface form of the query through token- or phrase-level changes intended to influence query interpretation. Examples include inserting misleading keywords, replacing entity-related terms, or modifying local phrasing around an ATT\&CK technique or mitigation entity. Semantic attacks instead modify the contextual meaning or instruction structure of the query while preserving the original cybersecurity reasoning target. For example, a semantic perturbation may introduce conflicting instructions or misleading contextual statements intended to redirect the model toward an incorrect ATT\&CK technique, mitigation, or graph relationship despite the underlying query objective. These two categories therefore capture complementary adversarial behaviours: lexical attacks target surface-level interpretation, whereas semantic attacks attempt to manipulate the higher-level reasoning intent of the generated query.

Category-specific robustness was measured using Attack Success Rate (ASR). GRICS achieved an ASR of approximately 68.3\% under lexical adversarial perturbations and 74.9\% under semantic adversarial perturbations, yielding an aggregate ASR of 71.6\% across the complete adversarial evaluation set. The higher ASR observed under semantic perturbations indicates that context-level manipulation presents a greater challenge to the reasoning pipeline than surface-level lexical modification. Tokens per Query (TPQ) was additionally used to characterize the amount of adversarial content required during the aggregate attack evaluation.

For broader context, Table~\ref{tab:asr_comparison} reports the robustness results of GRICS alongside representative PoisonRAG and GRAPPOISON results reproduced from \cite{liang2025graphrag}. Because these approaches were evaluated using different knowledge graphs, graph-construction procedures, adversarial implementations, and evaluation protocols, the comparison is intended to provide qualitative context rather than a controlled head-to-head assessment.
Although the robustness evaluation of GRICS was restricted to an ATT\&CK-oriented reasoning scope to better align with prior studies, the underlying BRIDG-ICS knowledge graph differs from the knowledge graphs used by PoisonRAG and GRAPPOISON in terms of ontology design, graph construction, and relational structure. Consequently, the reported ASR and TPQ values should be interpreted as providing qualitative insight into the robustness characteristics of graph-grounded retrieval systems rather than establishing direct quantitative superiority. A controlled comparison using identical knowledge graphs, attack settings, and evaluation protocols remains an important direction for future work.

\begin{table}[h!]
\centering
\caption{Qualitative comparison of adversarial robustness. Results for PoisonRAG and GRAPPOISON are reproduced from \cite{liang2025graphrag} and were obtained under different knowledge graphs, reasoning tasks, and attack settings.}
\label{tab:asr_comparison}
\begin{tabular}{l c c}
\hline
\textbf{Method} & \textbf{ASR (\%)} & \textbf{TPQ} \\
\hline
PoisonRAG    & 63.2  & 184.50 \\
GRAPPOISON   & 96.9  & 103.80 \\
GRICS        & 71.6  &  96.72 \\
\hline
\end{tabular}
\end{table}

\subsubsection{Robustness to Noisy Analyst Inputs}
\label{subsubsec:noisy_input}

In addition to intentional adversarial manipulation, practical cyber-threat intelligence systems must tolerate non-malicious imperfections in analyst queries. Such inputs may contain spelling errors, incomplete grammatical structures, or inconsistent formatting of cybersecurity identifiers while preserving the same underlying analytical intent. The noisy-input analysis therefore considers two settings: natural-language noise, combining typographical and grammatical perturbations, and entity-format noise. Unlike the adversarial perturbations evaluated in the previous subsection, these modifications are not intended to redirect the reasoning process toward an incorrect answer, but instead simulate realistic imperfections that may occur during analyst interaction.

For typographical and grammatical noise, the CTI-RCM 2024 benchmark was selected because its vulnerability descriptions contain substantial natural-language context, making it suitable for evaluating robustness when the linguistic formulation of a query is degraded. The perturbations introduce spelling errors, character-level modifications, omitted words, and grammatically incomplete or irregular sentence structures while preserving the original cybersecurity reasoning target. For example, a correctly formulated vulnerability-related query may be modified through misspelled security terminology or fragmented grammatical structure without changing the expected CWE classification.

Table~\ref{tab:noisy_input_robustness} reports the performance of the evaluated configurations under these combined natural-language perturbations. Base KG-RAG achieved an accuracy of 46.3\%, and the addition of fine-tuning alone resulted in the same accuracy of 46.3\%. This behaviour can be explained by the fact that fine-tuning improves the model's ability to generate ontology-compliant Cypher queries, but it does not correct incorrect or corrupted information contained in the input itself. Consequently, when typographical or grammatical perturbations distort the entity or relation expressed in the query, the generated Cypher statement may still be structurally valid while referring to an entity or pattern that does not exist in the knowledge graph, resulting in an empty or incorrect retrieval.

KG-RAG+EF achieved an accuracy of 82.7\%, while Full GRICS achieved a similar accuracy of 83.8\%. The substantial improvement over the configurations without embedding fallback indicates that semantic graph-anchor recovery is the primary mechanism responsible for robustness under noisy-input conditions. When typographical or grammatical perturbations prevent the initial symbolic query from reliably identifying the intended graph entity, the embedding mechanism can recover a semantically related anchor, after which an ontology-compliant Cypher query is regenerated and symbolic retrieval resumes. The small performance difference between KG-RAG+EF and Full GRICS is associated with cases involving omitted words or fragmented phrasing, where the fine-tuned Cypher-LLM may generate a more suitable alternative interpretation or select a more informative query phrase from the remaining context. In some instances, this can help Full GRICS recover a valid symbolic query that KG-RAG+EF does not produce. However, because this effect is limited and may depend on the model's generation behaviour, both configurations exhibit broadly comparable robustness, with embedding-assisted recovery remaining the dominant mechanism under noisy-input conditions.

\begin{table}[h!]
\centering
\caption{Robustness to combined typographical and grammatical noise on the CTI-RCM 2024 benchmark.}
\label{tab:noisy_input_robustness}
\begin{tabular}{lc}
\hline
\textbf{Configuration} & \textbf{Accuracy (\%)} \\
\hline
Base KG-RAG     & 46.3 \\
KG-RAG + FT     & 46.3 \\
KG-RAG + EF     & 82.7 \\
Full GRICS      & 83.8 \\
\hline
\end{tabular}
\end{table}

Meanwhile, entity-format noise was evaluated separately using the one-hop question set employed in both the runtime analysis (Section~\ref{subsec:runtime}) and the explainability and traceability analysis (Section~\ref{subsec:explainability_traceability}). This setting examines cases in which a cybersecurity identifier or entity name is expressed using a non-canonical format, including omitted separators, spacing variations, capitalization differences, or minor textual errors.

For entity-format perturbations, Base KG-RAG and KG-RAG+FT achieved 0\% successful resolution, whereas KG-RAG+EF and Full GRICS correctly resolved 100\% of the evaluated one- to three-hop queries when the intended identifier remained recoverable through semantic similarity. This difference arises because the entity identifier frequently serves as the principal matching term in the Cypher \texttt{WHERE} clause. When the identifier is malformed or expressed in a non-canonical format, Base KG-RAG and KG-RAG+FT cannot recover the corresponding graph entity and therefore fail to return relevant results. In contrast, configurations equipped with embedding fallback can identify the semantically closest graph entity, use it as an anchor, and regenerate an ontology-compliant Cypher query before symbolic execution resumes. The recovered embedding candidate is therefore used only to reconstruct the query and is not accepted directly as final evidence. For reasoning paths beyond three hops, the increased relational complexity may lead to incomplete or hallucinated query structures; this behaviour is examined further in Section~\ref{subsec:explainability_traceability}.

A different behaviour occurs when the numerical component of an identifier is modified. Changing the numeric portion of a CVE, CWE, CAPEC, or similar identifier may refer to a different valid cybersecurity entity rather than merely representing formatting noise. In this situation, GRICS does not assume that the altered identifier corresponds to the originally intended entity and may therefore return information associated with the identifier actually supplied in the query. This distinction prevents semantic fallback from incorrectly overriding potentially valid cybersecurity identifiers. These findings confirm that embedding-assisted recovery is the main mechanism supporting GRICS robustness to noisy natural-language and non-canonical entity inputs.

\subsubsection{Uncertainty-Aware Graph Reasoning}
\label{subsubsec:incomplete_graph}

In practice, complete deterministic mappings are not always available across the vulnerability-to-MITRE ATT\&CK reasoning chain. For example, not every CVE is associated with a CWE, not every CWE is linked to a CAPEC attack pattern, and not every CAPEC entry has a complete mapping to a MITRE ATT\&CK technique. As a result, a multi-hop reasoning path may become incomplete even though other potentially relevant associations are available.

The BRIDG-ICS ontology addresses this limitation by explicitly preserving candidate associations through relationship types such as \texttt{HAS\_POSSIBLE\_[NODE\_NAME]}~\cite{nandiya2025bridgicsaigroundedknowledgegraphs}. These relations represent potentially relevant mappings when a definitive relationship is unavailable. In this way, incomplete mappings are handled within the ontology itself rather than requiring the reasoning framework to infer unsupported relationships.

GRICS incorporates these uncertainty-aware relationships directly into its multi-hop reasoning process. When a query traverses the CVE--CWE--CAPEC--MITRE ATT\&CK chain, the generated Cypher queries can retrieve both definitive relationships and corresponding \texttt{HAS\_POSSIBLE\_*} relationships when they are represented in the graph. For example, if a CVE does not have a definitive CWE mapping but is connected through \texttt{HAS\_POSSIBLE\_CWE}, the associated possible CWE entities can still be returned as part of the graph-grounded evidence. Because GRICS generates multiple ontology-compliant Cypher candidates, different valid relationship patterns can be explored within the same analyst request.

The embedding-fallback mechanism further supports this process when the initial query cannot reliably identify the intended graph anchor. Once a relevant graph entity is recovered, GRICS regenerates an ontology-compliant Cypher query and retrieves the available direct and possible relationships associated with that entity. Importantly, the embedding mechanism is used only to recover a relevant graph anchor; it does not create or infer a missing relationship. The final evidence remains restricted to relationships explicitly represented in BRIDG-ICS. Robustness to incomplete mappings is therefore supported through the framework's ability to preserve and expose ontology-defined possible relationships during normal multi-hop retrieval. These possible associations remain explicitly identified by their relationship types and are not presented as confirmed mappings.

The practical purpose of this mechanism is to support analysts when a complete reasoning path is unavailable. Rather than terminating the investigation at the first missing deterministic mapping or requiring the analyst to restart the analysis from the original vulnerability, GRICS can expose the available candidate relationships and related entities. This allows the analyst to continue investigating relevant weaknesses, attack patterns, and MITRE ATT\&CK techniques while retaining visibility of which relationships are confirmed and which remain possible.

\subsection{Runtime Performance Analysis}
\label{subsec:runtime}

In addition to reasoning accuracy, the practical deployment of GRICS depends on its computational efficiency for real-world cyber threat investigations. Runtime performance was evaluated using 100 representative queries sampled from the evaluation datasets, comprising 20 entity lookup queries, 20 one-hop graph traversal queries, 40 multi-hop reasoning queries (covering both two-to-three-hop and four-to-five-hop reasoning), and 20 CTI-Benchmark queries.

The runtime analysis separately measures the latency of the three principal stages of the inference pipeline: (i) Cypher query generation, (ii) symbolic graph retrieval, and (iii) natural-language response synthesis. Timing measurements were performed using Python's \texttt{time.perf\_counter()} after all models had been loaded into GPU memory and initialized. Consequently, the reported latency represents steady-state online inference and excludes one-time model loading and initialization overhead.

Table~\ref{tab:runtime} summarizes the average latency for Cypher generation and symbolic graph retrieval across representative cybersecurity reasoning tasks. Entity lookup exhibits the lowest latency because queries are anchored to indexed entities, enabling efficient Cypher generation and localized graph traversal. As reasoning depth increases, both Cypher generation and symbolic retrieval require additional processing time due to increasingly complex graph traversal. Attack path analysis incurs the highest retrieval latency because substantially larger graph neighbourhoods must be explored.

In practice, runtime is dominated by the two LLM inference stages rather than graph retrieval. Depending on query complexity, Cypher generation requires between 0.86 and 2.58 seconds, while response synthesis requires approximately 1.46 seconds on average. Symbolic graph retrieval contributes comparatively little overhead for most query types, and the embedding-based fallback mechanism introduces only minimal additional latency because semantic similarity search is performed over precomputed node embeddings and is activated only when symbolic retrieval fails.

Although the current evaluation was conducted on the BRIDG-ICS knowledge graph, the architecture is designed to support larger industrial knowledge graphs, although its performance at substantially greater scale remains to be empirically validated. Symbolic retrieval performs localized graph traversal anchored by identified entities rather than exhaustive graph exploration, making query execution dependent primarily on reasoning depth and the retrieved subgraph rather than the total graph size. Furthermore, node embeddings are generated offline and reused during inference, allowing embedding-based retrieval to scale independently of LLM inference. Future deployments on million-scale knowledge graphs could further benefit from approximate nearest-neighbour indexing and distributed graph databases.

Table~\ref{tab:memory} summarizes the steady-state runtime memory utilisation after model initialization. The majority of GPU memory is occupied by the fine-tuned Llama-3.1-8B model, whereas the Neo4j knowledge graph and precomputed node embeddings reside in host memory. Since node embeddings are generated offline, embedding-based retrieval introduces only a modest memory overhead while enabling efficient semantic retrieval.

GRICS introduces an additional LLM inference stage for \textbf{\textit{ontology-aware Cypher generation}} before response synthesis. This increases inference cost and latency, but the additional computation enables the analyst query to be converted into an executable symbolic graph query prior to answer generation. The resulting trade-off is therefore between lower computational overhead and stronger ontology compliance, graph-grounded verification, and evidence traceability.

\begin{table}[t]
\centering
\footnotesize
\caption{Average latency across representative cybersecurity reasoning tasks.}
\label{tab:runtime}

\begin{tabularx}{\columnwidth}{>{\raggedright\arraybackslash}Xcc}
\toprule
\textbf{Query Category} &
\textbf{Cypher (s)} &
\textbf{Query Exec. (s)} \\
\midrule
Entity lookup & 0.86 & 0.82 \\
One-hop reasoning & 1.39 & 0.67 \\
Multi-hop (2--3 hops) & 1.61 & 0.96 \\
Multi-hop (4--5 hops) & 1.94 & 1.42 \\
CTI-Benchmark & 2.32 & 1.87 \\
Attack path analysis & 2.58 & 6.42 \\
LLM response synthesis & 1.46 & -- \\
\bottomrule
\end{tabularx}

\end{table}
\begin{table}[t]
\centering
\caption{Runtime memory utilisation of GRICS.}
\label{tab:memory}
\begin{tabular}{lc}
\toprule
\textbf{Component} & \textbf{Runtime Memory} \\
\midrule
Llama-3.1-8B (4-bit) & $\sim$6.3 GB GPU \\
Embedding model (MiniLM-L6-v2) & $<$100 MB \\
Neo4j knowledge graph & 1.9 GB Host RAM  \\
Node embeddings (precomputed) & $\sim$45 MB RAM \\
\bottomrule
\end{tabular}
\end{table}

\subsubsection{Computational Complexity and Scalability}
\label{subsubsec:complexity_scalability}

The computational cost of GRICS can be considered across four principal components: Cypher generation, symbolic graph retrieval, embedding-assisted fallback retrieval, and natural-language response generation. Let $N=|\mathcal{V}|$ denote the number of nodes in the knowledge graph, $d$ the embedding dimensionality, and $h$ the reasoning depth of the graph query. The two LLM-based stages, Cypher generation and response synthesis, are primarily influenced by model size and input/output sequence length and remain the dominant contributors to inference latency, as reflected in the runtime measurements reported above.

For symbolic retrieval, GRICS does not perform exhaustive traversal of the complete knowledge graph for each request. Generated Cypher queries are anchored to identified cybersecurity entities and retrieve only the relationships required by the requested reasoning path. Consequently, practical graph-query cost is influenced mainly by the size and density of the local neighbourhood, relationship branching factor, and reasoning depth $h$, rather than by direct traversal of all $N$ graph nodes. As the knowledge graph grows, retrieval latency is therefore expected to depend on whether the additional entities and relationships increase the neighbourhood explored by a particular query. This behaviour is reflected in the current runtime results, where deeper multi-hop and attack-path queries exhibit greater retrieval latency because larger relational neighbourhoods are traversed.

The embedding-assisted fallback mechanism introduces a separate scalability consideration. Node embeddings are generated offline and reused during inference, avoiding repeated embedding computation during normal query processing. For $N$ graph entities represented by $d$-dimensional embeddings, direct similarity comparison has an approximate computational cost of $O(Nd)$. As the graph grows, more efficient vector-retrieval mechanisms may therefore be required to prevent semantic fallback latency from increasing proportionally with the number of embedded entities.

Embedding regeneration also introduces an offline computational cost as the knowledge graph evolves. A complete regeneration requires embedding the textual representation of each graph entity and therefore increases approximately with the number of entities being processed. However, routine graph updates do not necessarily require regeneration of the complete embedding collection; newly introduced or modified entities can be embedded separately while unchanged representations are retained. Consequently, embedding maintenance primarily affects offline graph-update operations rather than the latency of every analyst query.

Memory requirements similarly increase with knowledge-graph size. In the current implementation, the knowledge graph occupies approximately 1.9~GB of host memory and the precomputed node embeddings approximately 45~MB, while the dominant GPU-memory requirement is the 4-bit Llama-3.1-8B model at approximately 6.3~GB, as reported in Table~\ref{tab:memory}. For a fixed embedding dimensionality, embedding storage grows approximately linearly with the number of embedded entities. Knowledge-graph storage also increases with both the number of entities and relationships, with relationship density becoming particularly relevant for multi-hop retrieval.

These observations suggest that increasing graph scale primarily affects the graph and vector-retrieval layers, whereas the memory required by the LLM remains largely independent of the number of graph entities. The current evaluation provides the measured computational baseline for BRIDG-ICS; substantially larger industrial knowledge graphs would require further empirical evaluation to determine the practical latency and memory behaviour under enterprise-scale workloads.

\subsection{Explainability and Reasoning Traceability}
\label{subsec:explainability_traceability}

Explainability in GRICS is evaluated from both quantitative and operational perspectives. The quantitative evaluation measures whether generated Cypher queries remain grounded in the BRIDG-ICS ontology, comply with query-generation constraints, and preserve structural validity as relational complexity increases. The operational evaluation examines whether the resulting reasoning paths can be inspected and verified by security analysts.

GRICS supports traceability through explicit multi-hop reasoning chains grounded in verifiable knowledge-graph entities and relationships. Rather than producing unsupported conclusions, the framework exposes the generated Cypher queries, retrieved entities, and traversed relationships associated with each result. This design allows analysts to \textbf{\textit{inspect intermediate reasoning steps, verify the supporting evidence}}, and contextualize model outputs within operational Industry~5.0 security workflows.

\subsubsection{Quantitative Explainability Metrics}
\label{subsubsec:quantitative_explainability}

Quantitative explainability is evaluated using a structured query corpus derived from the KG-RAG question-answering dataset described in Section~\ref{subsec:dataset}. The corpus enables systematic analysis of structural validity and reasoning behaviour across multi-hop queries. In particular, the evaluation examines whether generated queries remain grounded in the ontology and comply with query-generation instructions as relational complexity increases.

Three log-derived indicators are used: Hallucination Rate (HR), Query Violation Rate (QVR), and Schema Consistency Rate (SCR). These metrics capture complementary aspects of model behaviour, including grounding reliability, instruction adherence, and ontology conformity.
HR, defined in Equation~\ref{eq:hallucination_rate}, measures the proportion of generated queries containing fabricated entities, relationships, properties, or identifiers that are unsupported by the BRIDG-ICS ontology:

\begin{equation}
    \mathrm{HR}
    =
    \frac{N_{\mathrm{hall}}}{N_{\mathrm{gen}}},
    \label{eq:hallucination_rate}
\end{equation}

where $N_{\mathrm{hall}}$ denotes the number of hallucinated queries and $N_{\mathrm{gen}}$ denotes the total number of generated queries.
Query Violation Rate (QVR), defined in Equation~\ref{eq:query_violation_rate}, measures the proportion of evaluation runs in which the model violates the three-query generation constraint:

\begin{equation}
    \mathrm{QVR}
    =
    \frac{N_{\mathrm{viol}}}{N_{\mathrm{runs}}},
    \label{eq:query_violation_rate}
\end{equation}

where $N_{\mathrm{viol}}$ is the number of runs that violate the query-generation constraint and $N_{\mathrm{runs}}$ is the total number of evaluation runs.
Schema Consistency Rate (SCR), defined in Equation~\ref{eq:schema_consistency_rate}, measures the proportion of generated queries that use valid ontology entities, relationship types, and properties:

\begin{equation}
    \mathrm{SCR}
    =
    \frac{N_{\mathrm{valid}}}{N_{\mathrm{gen}}},
    \label{eq:schema_consistency_rate}
\end{equation}

where $N_{\mathrm{valid}}$ denotes the number of ontology-aligned queries. Lower HR and QVR values, together with a higher SCR value, indicate stronger grounding, improved structural control, and greater explainability.

\begin{table}[t]
\centering
\caption{Quantitative explainability metrics across reasoning hop depth.}
\label{tab:explainability_metrics_full}
\begin{tabular}{l|ccc|ccc}
\hline
& \multicolumn{3}{c|}{\textbf{Baseline}}
& \multicolumn{3}{c}{\textbf{Fine-Tuned}} \\
\textbf{Hop}
& \textbf{HR}
& \textbf{QVR}
& \textbf{SCR}
& \textbf{HR}
& \textbf{QVR}
& \textbf{SCR} \\
\hline
1-Hop & 0.35 & 0.20 & 0.82 & 0.15 & 0.10 & 0.93 \\
2-Hop & 0.45 & 0.28 & 0.76 & 0.22 & 0.15 & 0.84 \\
3-Hop & 0.58 & 0.40 & 0.53 & 0.30 & 0.22 & 0.67 \\
4-Hop & 0.78 & 0.60 & 0.25 & 0.45 & 0.35 & 0.55 \\
5-Hop & 1.00 & 0.80 & 0.00 & 0.70 & 0.50 & 0.35 \\
\hline
\end{tabular}
\end{table}

Table~\ref{tab:explainability_metrics_full} summarizes explainability performance across one- to five-hop reasoning. Both models exhibit comparatively stable behaviour on shallow queries, while structural degradation becomes increasingly evident beyond three hops. As relational depth increases, HR and QVR increase while SCR decreases, reflecting the growing difficulty of maintaining valid graph structure, ontology alignment, and query-generation constraints.

Despite this degradation, the fine-tuned model consistently outperforms the baseline at every reasoning depth. For one-hop queries, fine-tuning reduces HR from 0.35 to 0.15 and increases SCR from 0.82 to 0.93. This indicates that fine-tuning improves structural reliability even when the required graph traversal is relatively simple.
The performance difference becomes more pronounced as relational complexity increases. At four hops, fine-tuning reduces HR from 0.78 to 0.45 and QVR from 0.60 to 0.35, while increasing SCR from 0.25 to 0.55. At five hops, the baseline exhibits complete structural failure, with an HR of 1.00 and an SCR of 0.00. In contrast, the fine-tuned model retains partial structural validity, obtaining an HR of 0.70 and an SCR of 0.35. Fine-tuning markedly improves ontology conformity, instruction adherence, and grounding reliability as relational complexity increases. While performance drops for both models at deeper hop levels, the fine-tuned model declines more slowly and retains more structurally interpretable queries.

\subsubsection{Reasoning Completeness Across Hop Depth}
\label{subsubsec:reasoning_completeness}

The effect of relational depth on reasoning completeness is analyzed across increasing hop complexity within the BRIDG-ICS ontology using the same 450-query evaluation corpus. A query is considered successful when the model generates a correct and executable Cypher query that retrieves the intended graph path.

\begin{figure*}[t]
    \centering

    \begin{subfigure}[t]{0.48\linewidth}
        \centering
        \includegraphics[width=\linewidth]
        {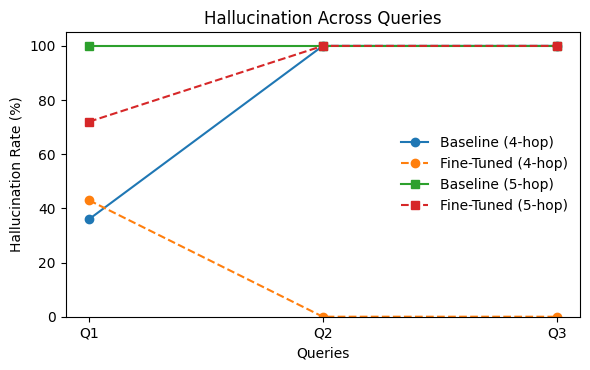}
        \caption{Multi-hop success rate across one- to five-hop queries originating from CVE entities.}
        \label{fig:multi-hop}
    \end{subfigure}
    \hfill
    \begin{subfigure}[t]{0.48\linewidth}
        \centering
        \includegraphics[width=\linewidth]
        {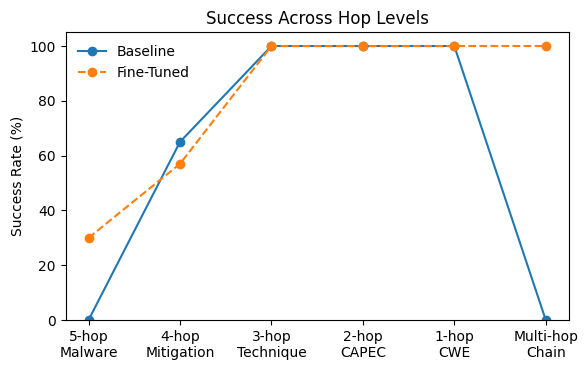}
        \caption{Hallucination rate across four- and five-hop queries.}
        \label{fig:halu}
    \end{subfigure}

    \caption{Reasoning completeness and structural stability under increasing multi-hop complexity.}
    \label{fig:reasoning_completeness}
\end{figure*}

In the baseline model, the three generated query alternatives frequently contain mismatched identifiers, invalid relationships, or repetitive query patterns. These errors result in incomplete or non-executable graph traversals. In contrast, the fine-tuned model produces more structurally coherent alternatives using ontology-defined relationships such as
\texttt{HAS\_POSSIBLE\_\allowbreak CWE} and
\texttt{HAS\_POSSIBLE\_\allowbreak TECHNIQUE}.
Consequently, alternative queries remain semantically grounded even when the first generated query is unsuccessful.
As shown in Figure~\ref{fig:multi-hop}, both models perform comparatively well on one- to three-hop queries, where the required relational chains remain relatively shallow. Structural difficulty becomes more apparent at four hops, where the model may need to reconstruct a complete relation chain such as

\begin{equation}
    \mathrm{CVE}
    \rightarrow
    \mathrm{CWE}
    \rightarrow
    \mathrm{CAPEC}
    \rightarrow
    \mathrm{Technique}.
    \label{eq:multi_hop_chain}
\end{equation}

At this depth, the baseline exhibits increasing instability, whereas the fine-tuned model maintains stronger performance by using domain-adapted representations and ontology-aware relationship patterns learned during fine-tuning. These patterns support more coherent reconstruction of multi-layer graph traversals, including paths that span both information technology and operational technology entities. The divergence is most pronounced at four- and five-hop depths, where relational complexity increases substantially. At five hops, the baseline frequently fails to reconstruct complete multi-layer paths. As illustrated in Figure~\ref{fig:halu}, the fine-tuned model recovers complete paths of four or more hops in approximately 62\% of cases, compared with 22\% for the baseline.

These findings indicate that shallow reasoning involving one to three hops remains manageable for both models. However, fine-tuning substantially improves \textbf{\textit{structural continuity and reasoning completeness}} for deeper four- and five-hop queries.

\subsubsection{Structural Stability and Retrieval behaviour}
\label{subsubsec:structural_stability}

\begin{figure*}[t]
    \centering

    \begin{subfigure}[t]{0.23\linewidth}
        \centering
        \includegraphics[width=\linewidth]
        {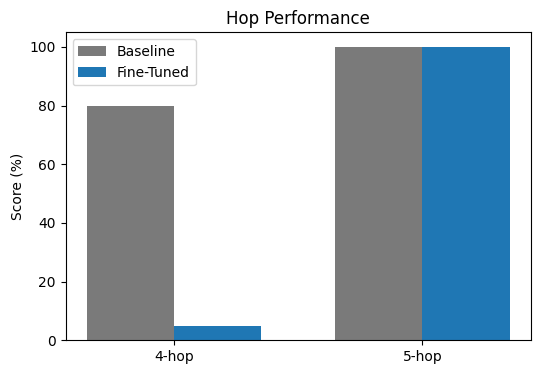}
        \caption{Query violation rate.}
        \label{fig:query_violation}
    \end{subfigure}
    \hfill
    \begin{subfigure}[t]{0.23\linewidth}
        \centering
        \includegraphics[width=\linewidth]
        {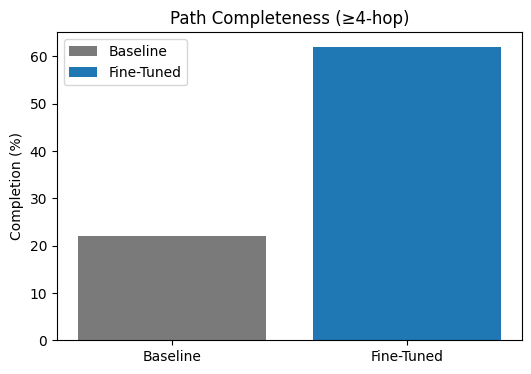}
        \caption{Complete-path recovery for queries requiring four or more hops.}
        \label{fig:reasoning_path}
    \end{subfigure}
    \hfill
    \begin{subfigure}[t]{0.23\linewidth}
        \centering
        \includegraphics[width=\linewidth]
        {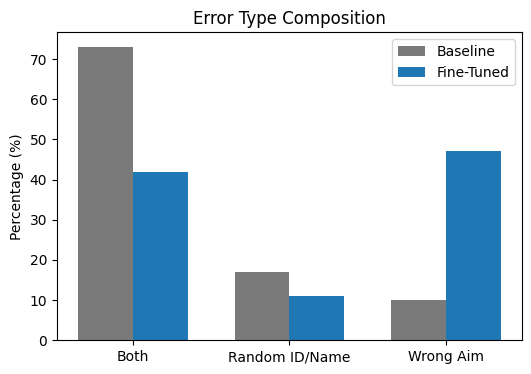}
        \caption{Retrieval-mode distribution.}
        \label{fig:ret_mode}
    \end{subfigure}
    \hfill
    \begin{subfigure}[t]{0.23\linewidth}
        \centering
        \includegraphics[width=\linewidth]
        {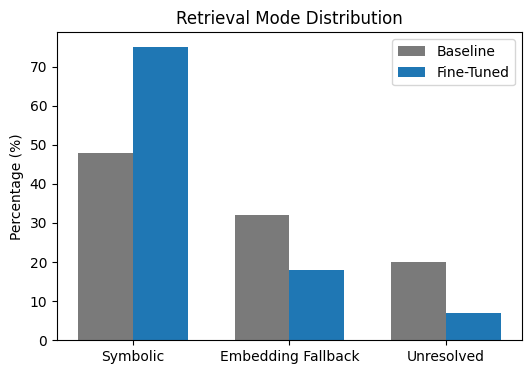}
        \caption{Error-type composition.}
        \label{fig:err_typed}
    \end{subfigure}

    \caption{Structural stability, retrieval behaviour, and error characteristics under multi-query generation across increasing hop depth.}
    \label{fig:stability_analysis}
\end{figure*}

Structural robustness under multi-query generation is evaluated through query-constraint adherence, reasoning completeness, retrieval behaviour, and error characteristics. These complementary analyses are summarised in Figure~\ref{fig:stability_analysis}.

Query Violation Rate provides an initial measure of structural control, as shown in Figure~\ref{fig:query_violation}. The baseline frequently exceeds the three-query constraint, particularly as relational complexity increases. In contrast, the fine-tuned model demonstrates stronger instruction adherence and more consistently generates query alternatives within the permitted boundary. This result indicates improved regulation of query generation under constrained reasoning conditions.

Reasoning completeness is further evaluated through complete-path recovery for multi-hop queries in Figure~\ref{fig:reasoning_path}, where the fine-tuned model achieves a substantially higher recovery rate than the baseline, indicating improved structural control and reasoning continuity. As shown in Figure~\ref{fig:ret_mode}, it also resolves more queries through direct symbolic graph traversal, whereas the baseline relies more heavily on embedding-based fallback and produces more unresolved cases. Direct symbolic traversal improves traceability by producing explicit and verifiable reasoning paths through the BRIDG-ICS ontology. Although fallback retrieval supports recovery when the initial Cypher query fails, it may introduce ambiguity when multiple graph nodes have similar semantic relevance; however, the retrieved anchors are used only to regenerate an ontology-compliant Cypher query before symbolic graph execution resumes.

Error-type composition provides further insight into model behaviour, as shown in Figure~\ref{fig:err_typed}. Baseline failures are predominantly compound errors that combine incorrect relationship targets with fabricated or mismatched node identifiers. The fine-tuned model produces fewer compound errors and generates outputs that remain more structurally interpretable, even when the retrieved path is incorrect. These results indicate that fine-tuning improves query-generation stability, adherence to structural constraints, reasoning continuity, and reliance on direct symbolic grounding. These improvements strengthen explainability by ensuring that a larger proportion of model outputs can be traced to explicit ontology entities, relationships, and executable graph paths.

Together, the quantitative metrics and behavioural analyses show that fine-tuning improves both GRICS performance and interpretability. Lower hallucination and query-violation rates indicate stronger grounding and instruction adherence, while higher schema consistency and complete-path recovery reflect better ontology alignment and reasoning continuity.
For analysts, these gains yield more transparent, verifiable outputs: generated Cypher queries reveal the relationships used in reasoning, retrieved graph paths provide explicit evidence, and error categories remain structurally interpretable. GRICS thus enables explainable cyber-threat reasoning by combining measurable structural reliability with evidence-linked reasoning traceability.

\subsection{Use Cases}

\begin{figure*}[!t]
    \centering
    \includegraphics[width=0.8\linewidth]{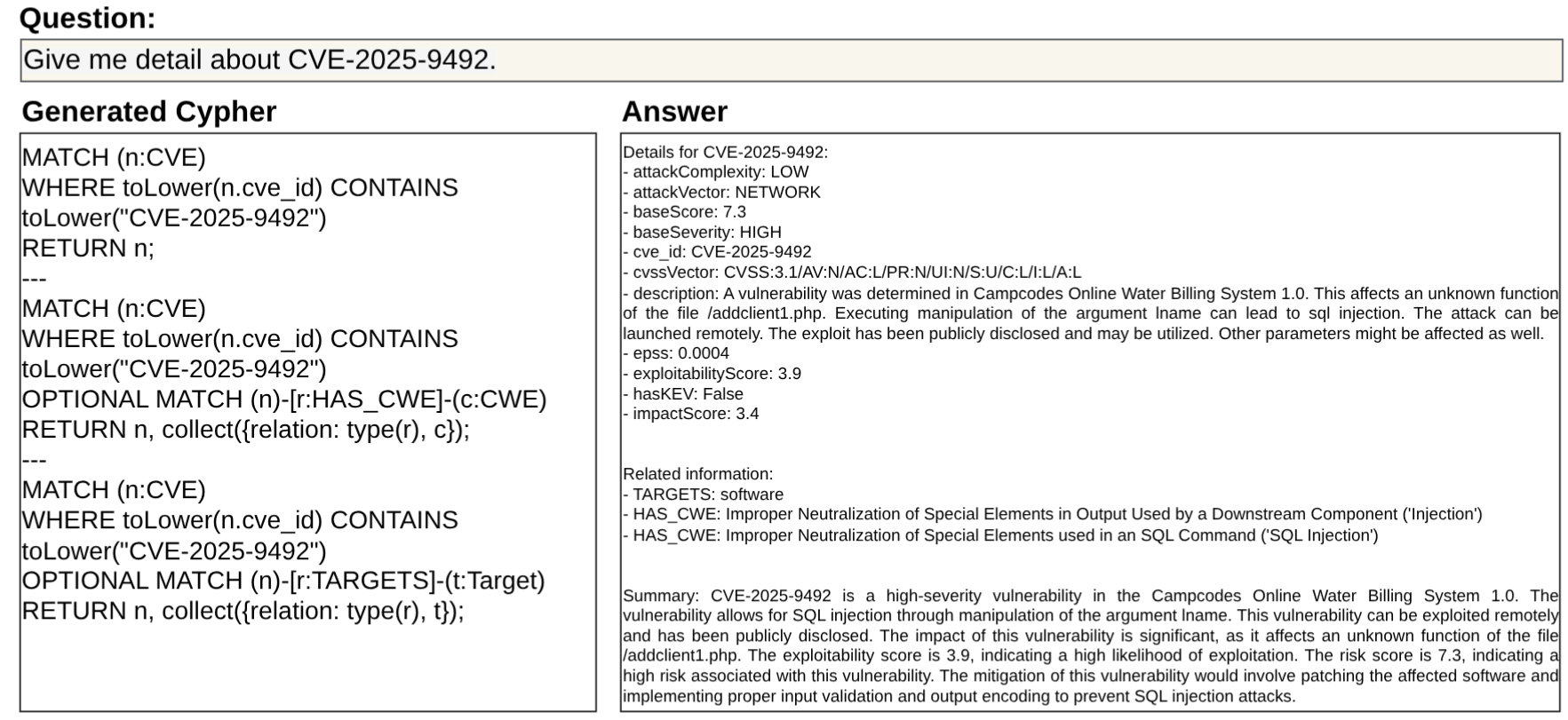}
    \caption{Example of vulnerability-centric analysis for a specific CVE.}
    \label{fig:cve_result}
\end{figure*}

This section presents representative use cases that demonstrate how GRICS performs multi-hop reasoning over heterogeneous cybersecurity knowledge graphs integrating MITRE ATT\&CK, NVD vulnerability data, CAPEC attack patterns, and industrial asset information. The selected examples are organized to reflect four core analytical capabilities of the framework: identifying attack paths across industrial assets, analysing vulnerability-specific evidence, attributing adversarial techniques through structured threat-intelligence relations, and deriving mitigations across interconnected semantic layers. In this way, the section shows how GRICS unifies asset-level, vulnerability-level, threat-level, and mitigation-level reasoning within a single graph-grounded framework for Industry~5.0 cybersecurity analysis.
\newline 
\textbf{Attack Path Reasoning.}
We first consider a multi-hop attack scenario within an industrial control system. Given the query \textit{``Show path between MQTT\_BROKER\_1 and SAFETY\_PLC\_2''}, GRICS automatically generates an executable Cypher query to retrieve valid graph paths between the specified assets. The resulting paths represent potential communication or attack routes formed by interconnected components, including brokers, runtime servers, conveyors, and PLCs. The path length reflects reasoning depth, capturing direct and indirect dependencies across network and operational layers. Retrieved paths depend on the underlying infrastructure; differences in topology, device configuration, segmentation policies, and Industry~5.0 architectures can yield different reasoning outcomes. These paths are then summarised in natural language, and context-aware mitigation recommendations are generated to support cyber--physical risk assessment.
\newline 
\textbf{Vulnerability-Centric Analysis.} 
Beyond asset connectivity, GRICS supports vulnerability-focused reasoning through structured graph retrieval. As shown in Figure~\ref{fig:cve_result}, a query for \textit{CVE-2025-9492} retrieves associated attributes such as CVSS severity scores, exploitability metrics, and linked CWE classifications. These elements are synthesized into a concise technical description of the vulnerability’s impact. To ensure correctness, precision, and traceability, the retrieved information is aligned with the official CVE record\footnote{\url{http://cve.org/CVERecord?id=CVE-2025-9492}}, demonstrating consistency with authoritative vulnerability data sources. Based on this validated knowledge, the system derives mitigation guidance, including patch deployment, input validation, and secure configuration practices.
\newline 
\textbf{Adversarial Technique Attribution.}
GRICS further enables adversarial behaviour analysis within the MITRE ATT\&CK domain. For the query \textit{``Which technique is used by group Cleaver via malware TinyZBot?''}, the framework performs multi-hop traversal across \textit{Group} Malware–Technique relationships in the knowledge graph. 
\begin{figure*}[!t]
    \centering
    \includegraphics[width=0.8\textwidth]{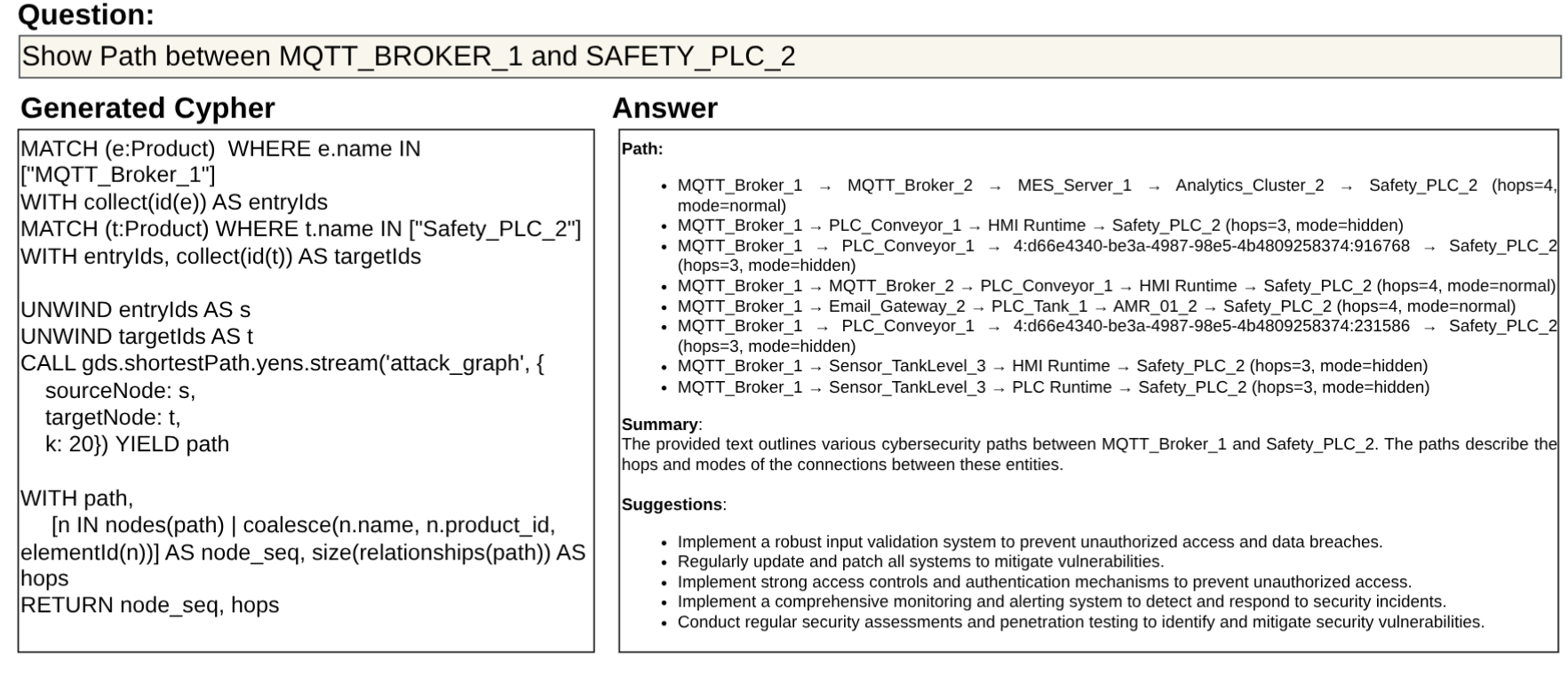}
    \caption{Example of multi-hop attack path reasoning between industrial assets.}
    \label{fig:attack_path}
\end{figure*}

\begin{figure*}[!t]
    \centering
    \includegraphics[width=0.8\linewidth]{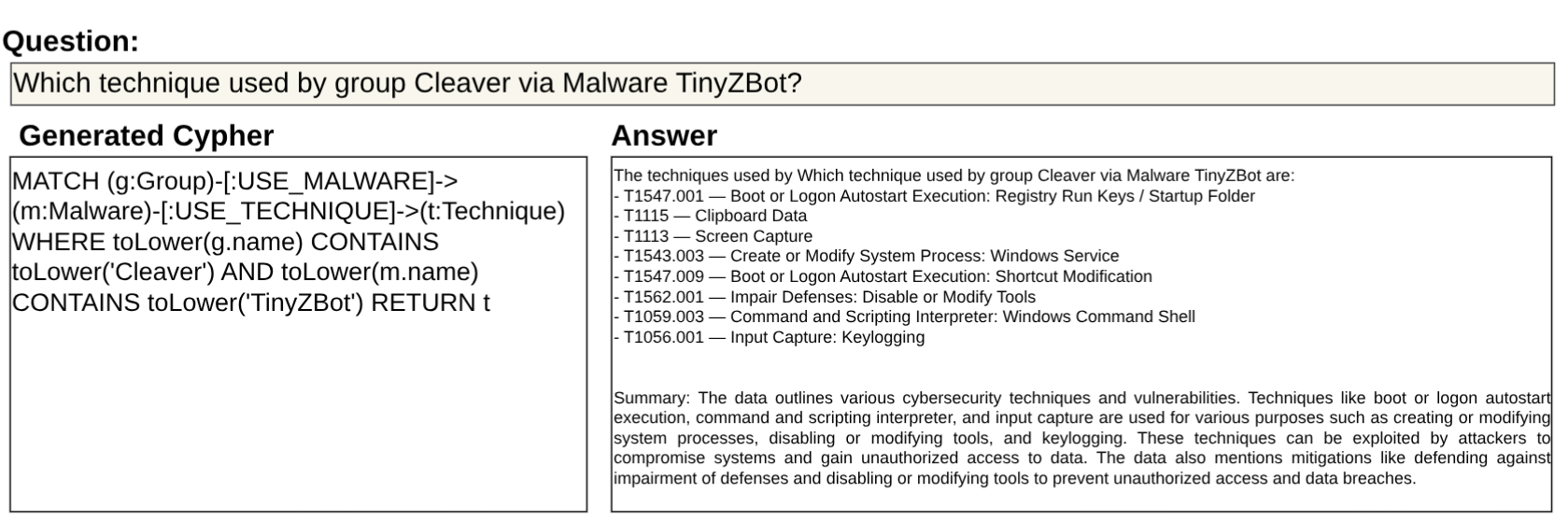}
    \caption{Adversarial technique attribution using multi-hop graph reasoning in GRICS. The framework traverses \textit{Group--Malware--Technique} relationships to identify relevant ATT\&CK techniques and generate structured explanations.}
    \label{fig:technique_result}
\end{figure*}

\begin{figure*}[!t]
    \centering
    \includegraphics[width=0.8\linewidth]{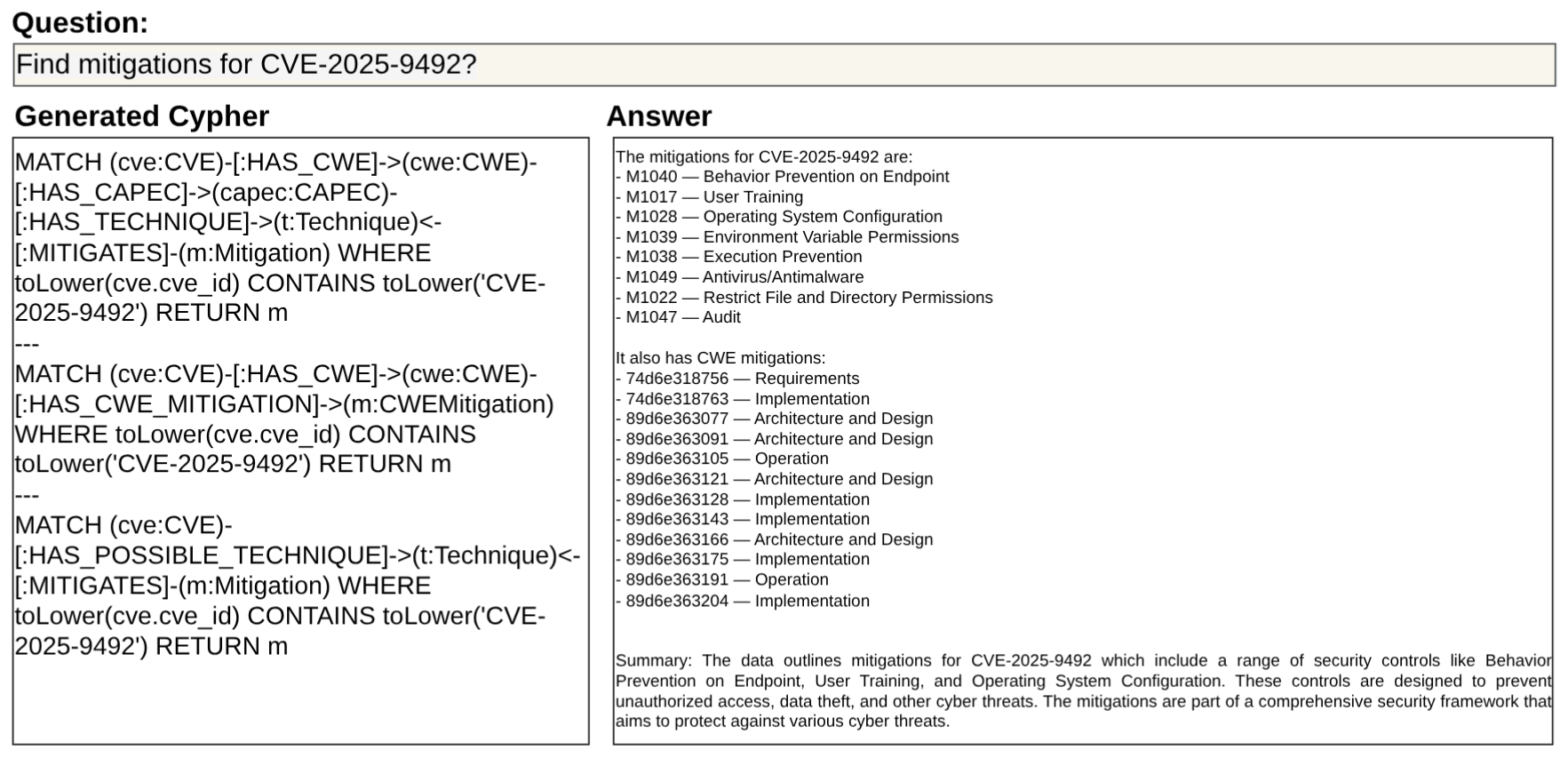}
    \caption{Example of automatically generated mitigation recommendations derived from multi-hop graph reasoning.}
    \label{fig:mitigation_result}
\end{figure*}

As illustrated in Figure~\ref{fig:technique_result}, the proposed framework identifies the relevant malware entity and its associated MITRE ATT\&CK techniques through structured graph traversal. The retrieved techniques are subsequently synthesized into a human-readable explanation, describing their operational intent and corresponding defensive considerations. The identified malware entity and its associated ATT\&CK techniques are consistent with the official MITRE ATT\&CK knowledge base, demonstrating the correctness of the graph-grounded reasoning process. This consistency is validated against the official MITRE ATT\&CK software entry for TinyZBot\footnote{\url{https://attack.mitre.org/software/S0004/}}.
\newline 
\textbf{Mitigation Derivation Across Abstraction Layers.}
GRICS derives solution-oriented mitigation strategies through multi-hop traversal across CVE, CWE, CAPEC, ATT\&CK, and mitigation entities. As shown in Figure~\ref{fig:mitigation_result}, the query \textit{``Find mitigations for CVE-2025-9492''} retrieves both ATT\&CK-level and CWE-level countermeasures, including behaviour prevention, operating system hardening, user awareness training, and secure design practices. All returned mitigations are consistent with ground-truth relationships encoded in the BRIDG-ICS knowledge graph. Fine-tuning further enables the model to infer additional plausible mitigation paths beyond explicitly linked edges, producing structurally valid recommendations through relational generalization. These inferred strategies remain semantically aligned with established countermeasures, indicating that fine-tuning supports controlled path expansion without introducing unsupported mitigation associations.

The presented use cases highlight the capability of GRICS to seamlessly integrate attack path analysis, vulnerability assessment, adversarial attribution, and mitigation planning within a unified graph-grounded reasoning framework. Through interpretable multi-hop inference and structured knowledge synthesis, the system minimizes manual correlation efforts and strengthens evidence-based cybersecurity decision-making in complex Industry~5.0 industrial ecosystems.

\subsubsection*{End-to-End Explainability Example}

\begin{figure*}[t]
    \centering
    \includegraphics[width=0.8\textwidth]{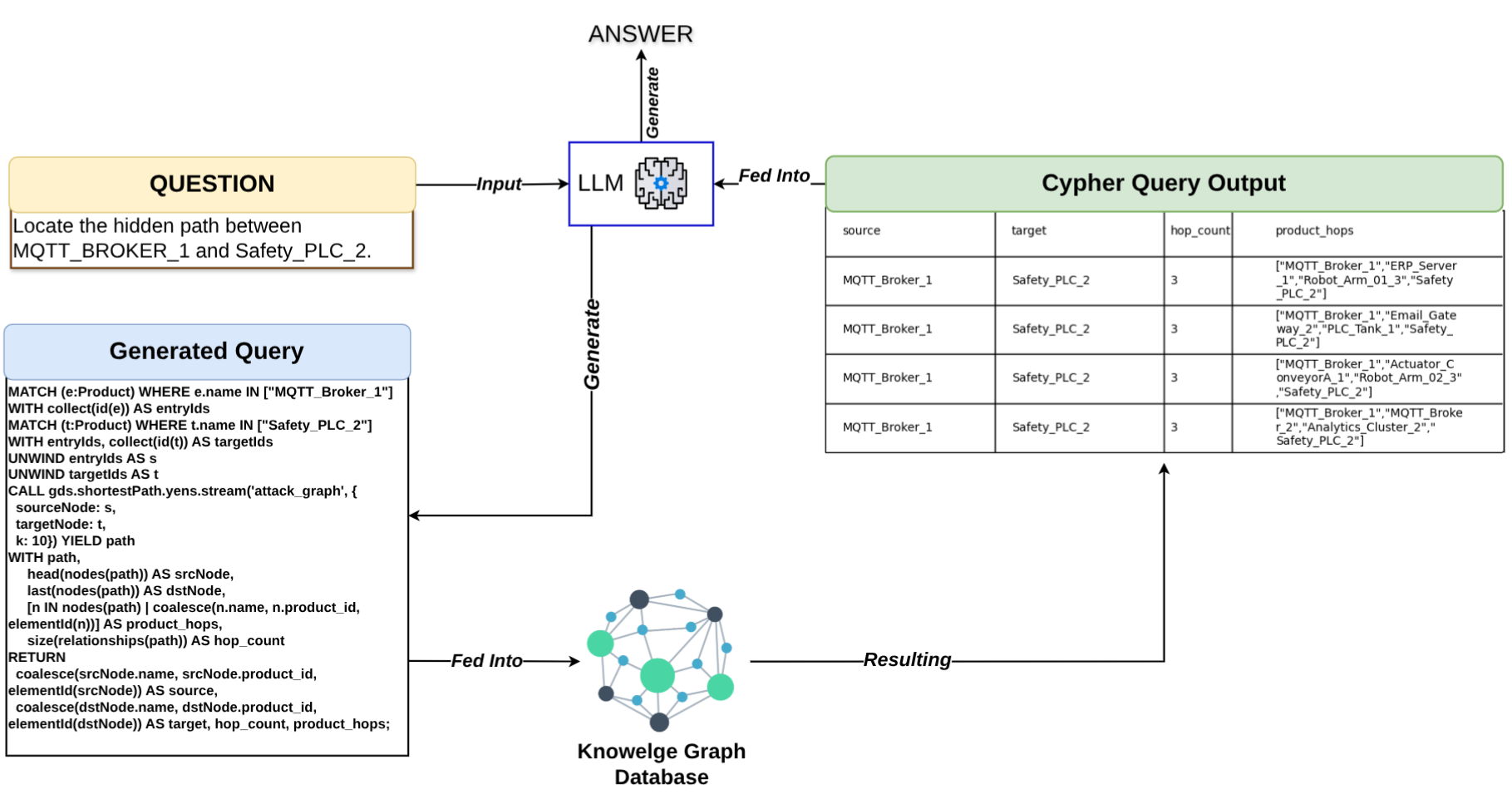}
    \caption{Explainability case study showing the analyst query, generated Cypher query, retrieved graph evidence, and final grounded response.}
    \label{fig:upd_pic}
\end{figure*}

Figure~\ref{fig:upd_pic} presents an end-to-end reasoning example illustrating the transparency of the proposed framework by exposing each stage of the graph-grounded inference pipeline. The example demonstrates how analysts can inspect the generated Cypher query, retrieved graph evidence, intermediate graph traversals, and the final grounded response, enabling every reasoning step to be traced and verified.

Unlike conventional LLM-based systems that directly generate responses, GRICS separates reasoning into two stages. The first LLM translates the analyst's natural-language query into an executable Cypher query, which is subsequently executed over the BRIDG-ICS knowledge graph. The retrieved graph evidence, including the discovered attack paths and intermediate entities, is then provided to a second LLM, which is not fine-tuned and is responsible solely for generating a human-readable summary of the retrieved evidence.

This graph-grounded design enables every generated response to be traced back to explicit knowledge graph evidence rather than opaque neural reasoning. Analysts can verify the generated Cypher query, inspect the returned graph entities and relationships, and confirm that the final response is fully supported by the retrieved evidence. Consequently, the reasoning process remains transparent, explainable, and auditable throughout the inference pipeline.  
Although this example is not a formal human-subject study, it illustrates the framework’s explainability artifacts that support analyst inspection, evidence verification, and informed cybersecurity decisions.

\section{Discussion}
\label{sec:discussions}
\subsection{Neuro-Symbolic Reasoning}

The experimental results demonstrate that GRICS improves cyber threat intelligence reasoning across both CTI-RCM and CTI-ATE tasks. These gains extend beyond predictive performance, reflecting a shift toward structured and semantically grounded reasoning enabled by the integration of fine-tuning and graph grounding. Unlike conventional text-based RAG approaches, which rely on pattern matching over unstructured data, GRICS leverages explicit relationships within the knowledge graph, preserving logical consistency across entities such as CVEs, CWEs, and ATT\&CK techniques. This is particularly critical in cybersecurity, where incorrect associations may propagate through downstream analysis.
Compared to existing Graph-RAG approaches such as GraphRAG~\cite{han2024graphrag} and KG2RAG~\cite{zhu2025}, which primarily rely on embedding-based or subgraph expansion strategies, GRICS introduces a tighter integration of symbolic query execution with neural retrieval. While these methods improve contextual relevance, they may still produce semantically plausible but structurally inconsistent relationships due to the absence of constraint-driven reasoning. In contrast, the Cypher-based symbolic retrieval in GRICS ensures that inferred relationships remain consistent with the underlying graph structure, improving reasoning fidelity and interpretability. Furthermore, relative to cybersecurity-focused approaches such as CyKG-RAG~\cite{cykg}, GRICS extends structured retrieval through fine-tuned query generation and hybrid fallback mechanisms, enabling more robust handling of ambiguous and compositional queries.

This structured reasoning also contributes to the robustness characteristics of GRICS. Under the ATT\&CK-oriented robustness evaluation, GRICS demonstrates resilience against adversarial prompt injection while preserving graph-consistent retrieval. Although it does not achieve the lowest Attack Success Rate (ASR), the results indicate that symbolic graph grounding provides a degree of protection against adversarial manipulation. Nevertheless, the evaluation also shows that graph grounding alone does not eliminate adversarial vulnerabilities, highlighting the need for more robust graph-aware defence mechanisms and standardised robustness evaluation protocols.

\subsection{Human--AI Collaborative Intelligence}
\label{subsec:human_ai_collaboration}

The findings indicate that GRICS is most appropriately positioned as a decision-support framework rather than an autonomous cybersecurity system. Its graph-grounded reasoning capabilities can reduce the effort required to correlate vulnerabilities, weaknesses, attack patterns, adversarial techniques, affected assets, and mitigations. However, the resulting recommendations should complement, rather than replace, expert judgement.

This collaborative role is particularly important in Industry~5.0 environments, where cybersecurity decisions may affect safety-critical assets, production continuity, and physical operations. By exposing generated Cypher queries, retrieved entities, and graph traversal paths, GRICS allows analysts to examine the evidence underlying each recommendation and determine whether it is consistent with the operational context. Human experts remain responsible for incorporating information that may not be represented in the knowledge graph, including asset configurations, organisational priorities, operational constraints, and acceptable risk levels.
The interaction between analysts and GRICS therefore combines complementary capabilities. The framework provides scalable knowledge correlation and structured multi-hop reasoning, while analysts contribute contextual understanding, domain expertise, and accountability. This division of responsibility supports more transparent and informed cybersecurity decision-making while limiting the risks associated with fully automated responses.

The explainability artifacts exposed by GRICS, including generated Cypher queries, retrieved entities, and explicit graph traversal paths, support analyst verification and accountable decision-making without transferring final authority to the automated system.

\subsection{Implications for Cybersecurity and Industry 5.0 Systems}

The findings have important implications for cybersecurity analysis in Industry~5.0 environments as follows: 
\newline 
\textbf{\textit{First}}, the ability to generate explainable multi-hop reasoning paths addresses a key limitation of existing AI-driven security systems, which often lack transparency. By exposing intermediate Cypher queries and graph traversal paths, GRICS enables analysts to verify reasoning steps, improving trust, auditability, and decision reliability in safety-critical contexts.
\newline 
\textbf{\textit{Second}}, the hybrid retrieval mechanism demonstrates that combining symbolic reasoning with embedding-based fallback provides a practical balance between precision and robustness. The embedding space analysis confirms stable semantic neighbourhoods, enabling reliable identification of relevant anchor nodes when symbolic query execution fails. This integration strengthens retrieval reliability by allowing semantic recovery while ensuring that the final evidence remains grounded through ontology-compliant symbolic graph execution.
\newline 
\textbf{\textit{Third}}, GRICS advances Human–AI collaboration by enabling analysts to interact with complex cyber-physical knowledge through natural language while maintaining traceability to underlying graph evidence. This capability is particularly relevant for Industry~5.0 systems, where tightly coupled IT/OT environments require interpretable and context-aware decision support.
\newline 
\textbf{\textit{Finally}}, the framework highlights the potential for integration with digital twin environments. By coupling graph-based reasoning with real-time or simulated system representations, GRICS could support dynamic risk assessment, attack path simulation, and proactive defence planning, enabling continuous monitoring and adaptive reasoning in cyber--physical systems.

\paragraph{\textbf{Scalability Considerations.}}

The current evaluation is conducted on the BRIDG-ICS knowledge graph and therefore provides an initial computational baseline for GRICS. For substantially larger enterprise knowledge graphs, scalability would depend on graph size, local relationship density, reasoning depth, embedding-search cost, and graph-update frequency. Because GRICS performs localized Cypher traversal rather than exhaustive graph exploration, retrieval latency is expected to be influenced more by the size and density of the queried neighbourhood than by total graph size alone, although deeper multi-hop queries may incur higher traversal cost.

Node embeddings are generated offline and reused during inference, so embedding regeneration is mainly required when graph entities are added or updated rather than for every query. Memory requirements would also increase with the number of graph entities, relationships, and stored embeddings, while the main GPU-memory requirement remains associated with the LLM. Performance on substantially larger industrial knowledge graphs has not yet been empirically validated and may require further optimisation or alternative architectural approaches for enterprise-scale deployment.

\paragraph{\textbf{Ontology Maintenance and Evolution.}}

Industrial cybersecurity knowledge evolves continuously as new vulnerabilities, attack techniques, assets, and threat relationships emerge. Although dynamic ontology evolution is not evaluated in the current implementation, GRICS can accommodate future BRIDG-ICS updates through periodic or continuous CTI ingestion. Newly collected CTI records could be normalised and mapped to the existing ontology before being synchronised with the Neo4j knowledge graph. New or modified graph entities would also require corresponding embedding updates to maintain consistency between symbolic and semantic retrieval.

More substantial ontology changes, such as the introduction of new entity classes or relationship types, would additionally require synchronisation with the ontology constraints used for Cypher generation. Maintaining alignment among the ontology, knowledge graph, embeddings, and query-generation instructions would therefore be an important consideration for future deployment in continuously evolving industrial environments.

\subsection{Limitations and Open Research Questions}
\paragraph{\textbf{Limitations.}}

The study is subject to the following limitations:

\begin{enumerate}
    \item \textbf{Reasoning and retrieval limitations.}
    Reasoning quality is sensitive to graph sparsity, ontology complexity, and increasing hop depth. Long or compositional queries may produce incomplete, redundant, or invalid Cypher statements, while the size of the BRIDG-ICS ontology introduces token overhead that may restrict the available context. The current evaluation also considers only one embedding model; therefore, the effects of alternative embedding approaches on retrieval quality and semantic separation remain uncertain. In addition, after an embedding-based graph anchor is identified, successful retrieval still depends on regenerating a valid ontology-compliant Cypher query. Predefined query patterns and overlapping candidate queries may further reduce flexibility and increase retrieval overhead.

    \item \textbf{Scalability and deployment limitations.}
    Although the current evaluation characterises runtime, memory utilisation, and computational complexity within the BRIDG-ICS environment, performance over substantially larger enterprise knowledge graphs has not yet been empirically validated. Larger graph structures may introduce additional retrieval latency, embedding-maintenance cost, memory requirements, and operational overhead depending on graph density and update frequency. Enterprise-scale deployment may therefore require additional optimisation strategies or alternative architectural approaches to maintain practical retrieval and reasoning performance as the knowledge graph grows.

    \item \textbf{Dual-LLM deployment considerations.} 
    The dual-LLM architecture introduces additional computational and deployment overhead compared with single-model RAG pipelines because separate inference stages are required for Cypher generation and response synthesis. This can increase inference cost, latency, memory requirements, and system-integration complexity, particularly when larger language models are employed. Although the current implementation can operate with locally deployed models, deployments that rely on externally hosted LLM services may additionally introduce API costs, network latency, service availability dependencies, and data-governance considerations. These trade-offs represent practical deployment limitations that should be considered when applying GRICS in resource-constrained or security-sensitive industrial environments.
    
    \item \textbf{Human-centred evaluation.} 
    Although GRICS provides graph-grounded explanations and traceable reasoning evidence, the current study does not evaluate these capabilities with cybersecurity practitioners. Future work could involve user studies with analysts performing representative threat-investigation tasks to assess explanation usefulness, trust, evidence verification, and decision-making efficiency. Such evaluation would provide a more direct assessment of the practical value of GRICS for human--AI collaboration in cybersecurity operations.

    \item \textbf{Cross-framework robustness comparison.}
    Although the robustness evaluation was restricted to ATT\&CK-oriented reasoning tasks to improve comparability with PoisonRAG and GRAPPOISON, differences remain in the underlying knowledge graphs, ontology structures, retrieval architectures, and adversarial settings. Consequently, the reported ASR and TPQ results provide qualitative insight rather than a fully controlled comparison, and direct quantitative superiority over these frameworks cannot be established.
\end{enumerate}

\paragraph{\textbf{Open Research Questions}}
The limitations observed in this study highlight several unresolved challenges in graph-grounded cyber threat intelligence. Reasoning over industrial knowledge graphs containing millions of entities requires scalable indexing, retrieval, and traversal mechanisms that preserve explainability. Future work should therefore investigate approximate nearest-neighbour indexing, distributed graph processing, and more efficient graph representations. Cybersecurity ontologies must also adapt to emerging threats, vulnerabilities, and relationships with limited manual intervention, motivating research into automated cyber threat intelligence ingestion, incremental graph updates, and ontology evolution. In addition, controlled robustness evaluations using identical knowledge graphs, attack settings, and evaluation protocols are required to support fair comparison with other Graph-RAG systems. Finally, systematic studies involving cybersecurity analysts are needed to assess usability, interpretability, cognitive workload, decision quality, and analyst trust in operational environments.

\section{Conclusion}

This paper presents GRICS, a Human–AI collaborative knowledge-graph framework for explainable threat reasoning in advanced manufacturing and Industry 5.0 environments. The study addresses a key limitation of existing RAG and Graph-RAG approaches for cybersecurity, which often struggle to maintain structurally valid multi-hop reasoning across tightly coupled IT and OT systems. The proposed framework addresses this challenge by unifying the BRIDG-ICS ontology, Cypher-based symbolic retrieval, controlled embedding fallback, and LLM answer synthesis in a single neuro-symbolic pipeline. Experiments show that it improves reasoning accuracy and usability. On CTI-RCM and CTI-ATE benchmarks, it consistently outperformed baseline LLMs, with the fine-tuned, graph-enhanced setup performing the best. The framework also increases retrieval transparency by resolving more queries via symbolic graph traversal, reducing reliance on approximate fallback, and generating traceable evidence chains. Robustness analysis shows that graph-grounded reasoning better preserves coherent multi-hop inference under adversarial and structurally complex queries. By exposing intermediate Cypher queries, graph evidence, and reasoning paths, the framework supports interpretable Human–AI collaboration, analyst trust, and informed decision-making in safety-critical settings.

Future work will investigate Symbolic Intelligence-driven digital twins, augmented with AI and ontological reasoning, alongside adaptive retrieval strategies and real-time cyber–physical threat monitoring. Combining graph-grounded reasoning with dynamic cyber-physical system representations offers a promising path for real-time risk assessment, attack simulation, and adaptive defence planning. Such advances would bridge static threat intelligence with evolving system states, positioning the framework as a foundation for next-generation, resilient, and explainable cybersecurity assistants in complex industrial environments.

This work demonstrates that graph-grounded neuro-symbolic reasoning provides an effective foundation for explainable cyber threat intelligence in Industry~5.0 environments. By combining symbolic graph retrieval with large language model reasoning, GRICS enables transparent, context-aware, and multi-hop threat analysis while maintaining traceability to structured cybersecurity knowledge. We believe this framework provides a promising foundation for future graph-grounded cybersecurity assistants supporting advanced manufacturing systems.


\section*{Declarations}

\noindent\textbf{Data Availability.}
The datasets used in this study are publicly accessible via the project’s GitHub repository:
\href{https://github.com/ahmadspm/Nuero-RAG-GRAPh-CPS-Intelligence-Modelling/tree/main/dataset}{\textit{Nuero-RAG-GRAPh-CPS-Intelligence-Modelling}}.

\vspace{0.3em}

\noindent\textbf{Funding.}
This work is supported by the Edith Cowan University, Australia Early and Mid-Career Research (EMCR) Grant.

\vspace{0.3em}

\noindent\textbf{Acknowledgements.}
We acknowledge the School of Science (Computing and Security Discipline) for providing access to the Industry~5.0 systems testbed and GPU resources used for LLM-driven knowledge graph enrichment and fine-tuning.

\vspace{0.3em}

\noindent\textbf{Generative AI Use.}
The authors acknowledge the use of OpenAI ChatGPT to assist with language refinement and grammatical review.
\vspace{0.3em}

\noindent\textbf{Competing Interests.}
The authors declare no competing interests.
\vspace{0.3em}

\section*{CRediT authorship contribution statement}
Writing – original draft: P.N., A.M.; Writing – review and editing: A.M., P.N., A.I., I.H.S., H.J.; Project administration: A.M.

\appendix

\bibliographystyle{spphys}
\bibliography{references}

\end{document}